\documentclass[aps,prd,floatfix,preprintnumbers,altaffilletter,superscriptaddress,reprint,longbibliography,twocolumn]{revtex4-1}

\usepackage{amssymb}
\usepackage{amsmath}
\usepackage{aas_macros}
\usepackage{hyperref}
\usepackage{color,units}
\usepackage{xspace}
\usepackage{subfigure}
\usepackage{comment}
\usepackage{xcolor}
\usepackage{bm}
\usepackage{dcolumn}
\usepackage{hyperref}
\usepackage{enumitem} 
\usepackage{array}
\usepackage{nicematrix}
\usepackage{hhline}
\usepackage{graphicx}
\usepackage{multirow}
\usepackage[normalem]{ulem}
\usepackage{soul}
\usepackage[math]{cellspace}
\usepackage{nicematrix} 
\usepackage{makecell}
\usepackage{acronym}
\usepackage{caption}

\begin{document}

\newacro{GW}{gravitational wave}
\newacro{BH}{black hole}
\newacro{BBH}{binary black hole}
\newacro{BNS}{binary neutron star}
\newacro{NSBH}{neutron star-black hole}
\newacro{LVK}{LIGO-Virgo-KAGRA Collaboration}
\newacro{GWTC-3}{the third LIGO--Virgo--KAGRA (LVK) gravitational-wave transient catalog}
\newacro{GWTC-2}{the second LIGO--Virgo--KAGRA (LVK) gravitational-wave transient catalog}
\newcommand{\default}{\textsc{Default}\xspace}
\newcommand{\exdefault}{\textsc{Extended}\xspace}
\newcommand{\nonid}{\textsc{NonIdentical}\xspace}
\newcommand{\isp}{\textsc{IsoSubPop}\xspace}
\newcommand{\gaussian}{\textsc{Gaussian}\xspace}

\newcommand{\tongh}[1]{\textcolor{red}{TH: #1}}
\newcommand{\sg}[1]{\textcolor{teal}{SG: #1}}
\title{The population properties of spinning black holes using \\ 
Gravitational-wave Transient Catalog 3}

\author{Hui Tong}
\email{hui.tong@monash.edu}

\author{Shanika Galaudage}

\author{Eric Thrane}
\affiliation{School of Physics and Astronomy, Monash University, VIC 3800, Australia}
\affiliation{OzGrav: The ARC Centre of Excellence for Gravitational-Wave Discovery, Clayton, VIC 3800, Australia}

\begin{abstract}
Binary black holes formed via different pathways are predicted to have distinct spin properties. 
Measuring these properties with gravitational waves provides an opportunity to unveil the origins of binary black holes. 
Recent work draws conflicting conclusions regarding the spin distribution observed by LIGO--Virgo--KAGRA (LVK).
Some analyses suggest that a fraction of the observed black-hole spin vectors are significantly misaligned (by $>90^\circ$) relative to the orbital angular momentum.
This has been interpreted to mean that some binaries in the LVK dataset are assembled dynamically in dense stellar environments.
Other analyses find support for a sub-population of binaries with negligible spin and no evidence for significantly misaligned spin---a result consistent with the field formation scenario.
In this work, we study the spin properties of binary black holes in the third LVK gravitational-wave transient catalog. 
We find that there is insufficient data to resolve the existence of a sub-population of binaries with negligible black-hole spin (the presence of this sub-population is supported by a modest Bayes factor of 1.7).
We find modest support for the existence of mergers with extreme spin tilt angles $> 90^\circ$ (the presence of extreme-tilt binaries is favored by a Bayes factor of 10.1).
Only one thing is clear based on gravitational-wave measurements of black hole spin: at least some of the LVK binaries formed in the field.
\textit{At most} $89\%$ of binaries are assembled dynamically (99\% credibility), though, the true branching fraction could be much lower, even negligible.
\end{abstract}

\maketitle

\section{Introduction}
The first detection of gravitational-wave events from a merger event of \ac{BBH} by LIGO--Virgo in 2015 \cite{PhysRevLett.116.061102} opened a new era of gravitational-wave astronomy. 
Since then, approximately 90 candidate gravitational waves from compact binary coalescences have been detected and recorded in \ac{GWTC-3} \cite{2021arXiv211103606T}. 
Most events are attributed to binary black hole (BBH) mergers with a handful of binary neutron star and neutron star + black hole mergers.
Other catalogues have also been produced by independent analysis using public data \cite{Nitz_2021,Olsen_2022}.
The LVK transient catalogs record the properties of each event including the component masses, spin vectors, and luminosity distance.
By studying the population properties of BBH systems, it is possible to infer how black holes form from massive stars and how they are assembled into merging binaries.

Binary black hole systems are thought to evolve via two main channels: either from the isolated evolution of massive binary stars, through a process known as the field scenario; or in star clusters, through a process known as the dynamical scenario \cite{Mapelli_2020}. 
Field binaries tend to have black-hole spins preferentially aligned with the orbital angular momentum due to tidal interactions.
On the other hand, the black-hole spin vectors in dynamically formed \ac{BBH} systems are expected to be distributed isotropically due to dynamical exchanges. These distinct predictions for black-hole spins provide a unique opportunity to study the fraction of current observed \ac{BBH} systems related to each channel. Inspired by this idea, many recent works (\cite{Farr_2011,Rodriguez_2016,Vitale_2017,Stevenson_2017,Farr_2017,Talbot_2017,Gerosa_2018,Franciolini:2022iaa,Callister_2021,Do_unequal,gwtc2_pop,2021_better_spin,LIGOScientific:2021psn,no_evidence,mrr,Biscoveanu_2022,Fishbach_2022}) seek to reveal the formation of binary black holes through the study of spin distribution in \ac{BBH} population observed by Advanced LIGO\citep{AdLIGO} and Virgo\citep{Virgo}, sometimes with contradictory conclusions.

The spin vector of each binary component is characterized by a spin magnitude $\chi_{1,2}$, a tilt angle $\theta_{1,2}$, and an azimuthal angle $\phi_{1,2}$.
Here the subscripts denote whether the parameter refers to the more massive (primary) or less massive (secondary) black hole.
Each angle is measured in a coordinate system with the $z$-axis aligned with the orbital angular momentum. 
Since black hole spin vectors can vary with time due to precession, it is useful to define an additional parameter, which is an approximate constant of motion. The effective inspiral spin $\chi_{\rm{eff}}$ \cite{Damour_2001,Ajith_2011}, 
\begin{equation}
\chi_{\rm{eff}}=\frac{\chi_1 \cos\theta_1+q\chi_2\cos\theta_2}{1+q} ,
\end{equation}
is a mass-weighted average of spin components projected along the orbital angular momentum. Here, $q=m_2/m_1$ is the mass ratio. 

Using data from LVK gravitational-wave transient catalog 2 (GWTC-2), Ref.~\citet{gwtc2_pop} found that $12\%$ to $44\%$ of \ac{BBH} systems merge with negative $\chi_{\rm{eff}}$, implying that a fairly large fraction of \ac{BBH} systems merge with significantly misaligned black hole spin vectors. 
This result was interpreted as evidence for dynamical mergers since it is difficult to produce such large misalignment angles through supernova kicks \cite{Stevenson_2022}.
However, this conclusion was challenged when Ref.~\citet{Roulet_2021} pointed out that the evidence for significantly misaligned spin vectors is likely due to model misspecification \citep{wmf}.
They argue that the evidence for $\chi_{\rm{eff}} < 0$ may actually comes from an unmodeled sub-population with $\chi_{\rm{eff}} = 0$. 
Ref.~\citet{2021_better_spin} follows up by exploring the possibility of a sharp feature near zero in the distribution of black hole spin magnitude. They find no clear evidence for significantly misaligned spin in \ac{GWTC-2} and report $29\%$ to $75\%$ \ac{BBH} systems merge with negligible spin ($90\%$ credibility). 

In the latest LVK analysis of \ac{GWTC-3} \citet{LIGOScientific:2021psn}, the LVK reiterates the presence of negatively aligned spins, with the minimum $\chi_{\rm{eff}} < 0$ at $88\%$ credibility, and less evidence for zero spin binaries. They reported $27-81\%$ of \ac{BBH}s are spinning. 
More detailed studies on the purported zero-spin sub-population have been made in Ref.~\citet{no_evidence}. They employed a series of variant models based on analyses in Refs.~\cite{LIGOScientific:2021psn} and found, although the possibility of a negligible-spin population is not precluded, an excess of zero-spin systems is not required by current data. Also, they show $\cos \theta$ confidently extends to negative values, with the lower truncation in the $\cos \theta$ distribution (i.e., hyper-parameter $z^\text{min}$ in this work) $\lesssim -0.35/-0.31$ ($95\%$ credibility) depending on the model.

Ref.~\citet{mrr} explores the idea of a zero-spin peak as well and find even less support than Ref.~\cite{2021_better_spin} for a sub-population of zero-spin mergers. 
They relax the assumption of identical distributions for $\chi_1$ and $\chi_2$, thus preserving the possibility of just one (non-) spinning \ac{BH} in binaries. They find that $< 46\%$ of primary black holes have negligible spin  and $<36$ \% of secondary black holes have negligible spin ($99\%$ credibility).
Only $\sim 1\%$ of mergers contain two black holes with negligible spins, a result which is seemingly inconsistent with Ref.~\cite{2021_better_spin}.

In this paper, we endeavour to help clarify some of the confusion surrounding the distribution of binary black hole spins.
To this end, we improve on the analysis from Ref.~\cite{2021_better_spin}: updating the analysis to include more events in \ac{GWTC-3}, documenting and correcting mistakes in the analysis code, and carrying out a more complete suite of model comparisons. The remainder of this paper is organized as follows. In Section \ref{sec:method} we describe our methodology, with special attention to improvements from \cite{2021_better_spin}. In Section \ref{sec:Results}, we present the results of our analyses. We conclude and discuss our findings in Section \ref{sec:Discussion and conclusion}.


\section{Methods}\label{sec:method}
We begin with the same set of 69 events as in Ref.~\cite{LIGOScientific:2021psn}, which are selected by requiring a false alarm rate $\text{FAR}\textless\unit[1]{yr^{-1}}$.
However, we flag two events, GW191109 and GW200129, as potentially problematic due to data quality issues. 
Reference \cite{Payne2022} have recently suggested that GW200129---an event had been hailed as an example of a precessing binary \cite{2021arXiv211103606T,Hannam2022}---may be an ordinary GW150914-like binary, which only \textit{appears} to be precessing due to a coincident glitch.
We therefore exclude GW200129 from our analysis entirely.
Meanwhile, unpublished (and currently inconclusive) work, leads us to question the reliability of inference results associated with GW191109---the event with the strongest signature of $\chi_\text{eff}<0$ in GWTC-3.
Since we are currently unsure of the reliability of GW191109, we carry out our analyses with and without GW191109.
Thus, we analyze 67-68 events depending on whether GW191109 is included. In the remainder of this paper, we mainly show results when GW191109 is excluded if there is not a significant difference between results of analyses with and without GW191109.

We adopt the \exdefault model from Ref.~\cite{2021_better_spin} as our baseline model, supplemented by some variants.
The \exdefault model is an extension of the \default spin model from the \ac{GWTC-3} population analysis \cite{LIGOScientific:2021psn}. 
It describes the distribution of component spin magnitudes and tilt angles (as opposed to the distribution of effective spin parameters).
In the \exdefault model, we assume the spin magnitude of each \ac{BH} contains a mixture of two sub-populations: spinning and non-spinning.
In this work, we split the \exdefault model into two versions:
\begin{widetext}
\begin{equation}
\begin{aligned}
    \pi(\chi_{1,2}|\alpha_\chi,\beta_\chi,\lambda_0) =  
    \begin{cases}
    (1-\lambda_0) \text{Beta}(\chi_{1}|\alpha,\beta)
    \text{Beta}(\chi_{2}|\alpha,\beta)
    +\lambda_0 \delta(\chi_1)\delta(\chi_2) & \textsc{Extended} \\
    (1-\lambda_0)
    \text{Beta}(\chi_{1}|\alpha_1,\beta_1)
    \text{Beta}(\chi_{2}|\alpha_2,\beta_2)
    +\lambda_0 \delta(\chi_1)\delta(\chi_2) & \nonid \\
    \end{cases} .
\end{aligned}
\end{equation}
\end{widetext}
Here, $\pi(\chi_{1,2}|...)$ is the prior distribution for the dimensionless spin magnitudes, which is conditioned on hyper-parameters $\alpha, \beta, \lambda_0$.
One sub-population of binaries contain spinning black holes with $\chi_1, \chi_2$ drawn from a non-singular Beta distribution with shape parameters $(\alpha,\beta)$ ($\alpha,\beta \geq 1 $) \footnote{For convenience, we re-parameterize the Beta distribution in terms of the spin magnitude mean $\mu_\chi$ and standard deviation $\sigma_\chi$ in our population analyses realization.}.
In the \exdefault variant, one set of hyper-parameters describes the distribution of both the primary spin $\chi_1$ and the secondary spin $\chi_2$.
In the \nonid variant, we use separate hyper-parameters to fit these two distributions.
The alternative sub-population is described by a delta function, which forces $\chi_1 = \chi_2 =  0$. 
As predicted by \cite{Fuller_2019}, BHs born from single stars may rotate very slowly, with $\chi \sim 10^{-2}$ due to efficient angular momentum transport.
It may follow that the majority of \ac{BBH} systems contain black holes with very low spins indistinguishable from zero using current observatories.
The mixing parameter $\lambda_0$ is the fraction of binaries with zero spin while $(1-\lambda_0)$ is the fraction with spin. However, due to the flexibility of Beta distribution model for spinning sub-population, it may also contribute to the negligible spin sub-model when the peak of Beta distribution $\lesssim 0.01$.

Following \cite{2021_better_spin}, we model the distribution of (the cosine of) black-hole spin tilts $z \equiv \cos\theta$ using another mixture model.
We introduce a few different variations:
\begin{widetext}
\begin{equation}
\begin{aligned}
    \pi(z_{1,2}&|\zeta,\sigma^t,z_\mathrm{min})
    =  
    \begin{cases}
    \zeta G_t{(z_1|\sigma^t,z^\text{min}})G_t(z_2|\sigma^t,z^\text{min})
    +(1-\zeta)\left(\frac{\Theta(z_1-z^\text{min})}{1-z^\text{min}}\right) \left(\frac{\Theta(z_2-z^\text{min})}{1-z^\text{min}}\right)  & \textsc{Extended} \\
    \zeta G_t(z_1|\sigma^t_1,z^{\text{min}}_1)G_t(z_2|\sigma^t_2,z^\text{min}_2)
    +(1-\zeta) \left(\frac{\Theta(z_1-z^\text{min}_1)}{1-z^\text{min}_1}\right) \left(\frac{\Theta(z_2-z^\text{min}_2)}{1-z^\text{min}_2}\right)  & \nonid \\
    \zeta G_t(z_1|\sigma^t,z^\text{min})G_t(z_2|\sigma^t,z^\text{min})
    + (1-\zeta)\left(\frac{1}{4}\right)  & \isp \\
    \zeta G_t(z_1|\sigma^t_1,z^\text{min}_1) G_t(z_2|\sigma^t_2,z^\text{min}_2) 
    + (1-\zeta)\left(\frac{1}{4}\right)  & \nonid \isp \\
    \end{cases} .
\end{aligned}
\end{equation}  
\end{widetext}
Here, $\pi(z_{1,2}|...)$ is the prior distribution for the cosine of the spin tilts, which is conditioned on hyper-parameters $\zeta, \sigma^t, z^\text{min}$.
$G_t(z|,\sigma^t,z^\text{min}$) is a truncated Gaussian distribution on the interval $[z^\text{min},1]$ with a peak at $z=1$ and width $\sigma^t$.
The factors of $\Theta(z-z^\text{min})/(1-z^\text{min})$ and $1/2$ are uniform distributions on the intervals $[z^\text{min}, 1]$ and $[-1,1]$ respectively.
The hyper-parameter $\zeta$ is the fraction of \textit{field-like} binaries, for which the black hole spin is preferentially aligned to the orbital angular momentum while $1-\zeta$ is the fraction of \textit{dynamical-like} binaries with quasi-isotropically\footnote{We use the phrase ``quasi-isotropically'' because the \exdefault and the \nonid model variants truncate the spin tilt distribution for dynamical-like binaries, and so the spin tilt distribution is not actually isotropic.} distributed spin.
We use the hyper-parameter $z^\text{min}$ to apply a maximum tilt angle.
Depending on the model variant, $z^\text{min}$ may apply to the entire population or just the sub-population of field-like binaries.

The \exdefault variant is the same as the one used in Ref.~\cite{2021_better_spin}.
The \nonid variant is the same as \exdefault except that the field-like primary and secondary spin distributions have different hyper-parameters $\sigma^t_1, \sigma^t_2, z^\text{min}_1, z^\text{min}_2$ while the \exdefault variant assumes that the primary and secondary spins have the same distribution with hyper-parameter $\sigma^t, z^\text{min}$.
This allows us to test whether the primary spin distribution and the secondary spin distribution are the same.

The \isp variant takes the \exdefault variant and moves the step function $\Theta(z-z^\text{min})$ so that it applies to only field-like binaries as opposed to all binaries.
While the \exdefault model is useful for testing whether there is support for \textit{any} binaries with $\chi_\text{eff}<0$, it does not allow for a realistic sub-population of dynamical mergers because the dynamical-like sub-population gets cut off at $z_\text{min}$.
The motivation for the \isp variant is to maintain the $z_\text{min}$ parameter, which seems to improve the fit of the \exdefault model \cite{2021_better_spin}, while allowing for a more realistic sub-population of dynamical binaries.
The \nonid \isp variant combines the \isp and \nonid variants.

\begin{table*}
\centering
 \begin{tabular}{|p{4cm}|p{9cm}|} 
 \hline
 Variant & Description \\
 \hline\hline
 \exdefault & The baseline model from Ref.~\cite{2021_better_spin}. No binaries merge with $z>z^\text{min}$ and $z_1, z_2$ are identically distributed. \\\hline
 \nonid & No binaries merge with $z>z^\text{min}$ and $z_1, z_2$ may have different distributions. \\\hline
 \isp & No field-like binaries merge with $z>z^\text{min}$, but dynamical-like binaries can; $z_1, z_2$ are identically distributed. \\\hline
 \nonid \isp & No field-like binaries merge with $z>z^\text{min}$, but dynamical-like binaries can; $z_1, z_2$ may have different distributions. \\\hline\hline
 \default & The LVK model from Ref.~\cite{gwtc2_pop}. There is no $z^\text{min}$ cutoff and $z_1, z_2$ are identically distributed. Does not include a sub-population of \ac{BBH} with zero spin. \\\hline
 \end{tabular}
 \caption{A summary of the model variants employed in this paper. The first four models allow for a sub-population with zero spin, parameterized by mixing fraction $\lambda_0$. However, each of these variants can be further subdivided into $\lambda_0=0$ (no zero-spin sub-population) and $\lambda_0>0$ (yes zero-spin sub-population) variants.}
 \label{tab:variants}
\end{table*}

Table~\ref{tab:variants} provides a summary of each variant. The full list of priors on various hyper-parameters is given in Table~\ref{tab:priors}. 
Following Refs.~\cite{gwtc2_pop,LIGOScientific:2021psn}, we adopt the \textsc{Power Law + Peak} model \cite{2018_colm} for the distribution of black-hole masses and a power-law distribution for redshift \cite{2018_maya}. 
We employ the selection effects treatment as used in Ref.~\cite{LIGOScientific:2021psn}.We make use of the same simulated injections used by Ref.~\cite{LIGOScientific:2021psn} to estimate the fraction of events in the Universe that would be detected for a particular population model. We neglect selection effects due to black-hole spin which are technically challenging to implement since there is a sharp feature in our black-hole spin model. We believe our results are still reliable since the selection effect from spin is relatively weak. Nonetheless, it is desirable to include selection effects in subsequent analyses using a dedicated injection set including a sub-population with negligible spin.
We analyze LVK samples from the \ac{GWTC-3} Parameter data release~\footnote{\url{https://zenodo.org/record/5546663}}.
We employ \texttt{GWPopulation} \cite{2019PhRvD.100d3030T} to perform hierarchical Bayesian inference, which utilizes \texttt{Bilby} \cite{2019_bilby,2020_bilby}.
\texttt{GWPopulation}  employs``recycling'' to evaluate marginalisation integrals with importance sampling  \citep{2019_Bayesian}.
In order for this method to be reliable, each likelihood evaluation requires a reasonably large number of effective samples. 
It can be challenging to recycle samples when using models with sharp features such as the sharp peak at $\chi=0$ in our distributions of black-hole spin.
Thus, to avoid undersampling, we supplement the LVK samples using purpose-built, zero-spin samples, which enable us to resolve the existence of a sharp $\chi=0$ feature.
We update the zero-spin samples used in Ref.~\cite{2021_better_spin}, which used \textsc{IMRPhenomD}, with the LVK ``preferred'' waveform. 
This is an improvement over Ref.~\cite{2021_better_spin} since we eliminate a possible source of bias arising from inconsistent use of waveforms for $\chi>0$ and $\chi=0$ sub-populations.
Our new samples are obtained using \texttt{BILBY} \cite{2019_bilby,2020_bilby} using the \textsc{IMRPhenomXPHM} waveform \cite{2021_IMRPhenomXPHM}, which incorporates higher-order modes. 

Additionally, we fix a mistake in Ref.~\cite{2021_better_spin} pointed out in Ref.~\cite{no_evidence}.
The authors of that work point out that the (spin / no-spin) Bayes factor for $\rm{GW}190408\_181802$ used in Ref.~\cite{2021_better_spin} is incorrect by two orders of magnitude, which leads to biased inferences about zero-spin binaries. 
Recalculating this using \textsc{IMRPhenomXPHM}, we obtain a (spin / no-spin) Bayes factor of $\mathcal{B} \sim 2.71$.
This result is more nearly consistent with the value of $\mathcal{B} \sim 1.6$ calculated using the Savage-Dickey density ratio formula in Ref.~\cite{no_evidence}. 
We suspect that Ref.~\cite{2021_better_spin} performed this calculation using slightly different strain data for the spinning and non-spinning analysis---possibly due to different de-glitching processes, which would still lead to reasonable posterior distributions, but an incorrect Bayes factor. 

Before moving on to the results, we summarize the main differences between this work and Ref.~\cite{2021_better_spin}:
\begin{itemize}
    \item We update the analysis to use data from GWTC-3.
    \item We consider additional model variations, allowing for nonidentical distributions of primary and secondary spin and also different interpretations of the $z_\text{min}$ parameter.
    \item We employ a new set of zero-spin samples, which uses the same waveforms as the official LVK samples.
    \item We correct a mistake identified by Ref.~\cite{no_evidence}, which biases the inferences in Ref.~\cite{2021_better_spin}.
    Erratum changes to Ref.~\cite{2021_better_spin} are described in footnote~\footnote{The updated version of Ref.~\cite{2021_better_spin} corrects for another bug, which resulted in the non-spinning posteriors to be given more weight where the fiducial prior per event was 4 times larger than it should have been. However, the update to \cite{2021_better_spin} does not address the suspect samples for $\rm{GW}190408\_181802$, which we fix here.}.
\end{itemize}

\section{Results}\label{sec:Results}
\subsection{Model selection}
We carry out population inference using the model variants summarized in Table~\ref{tab:variants}.
Our findings---excluding GW191109---are summarized in Table~\ref{tab:BF2}.
The table shows both Bayes factors and maximum likelihood ratios in order to separate out how the Bayes factor is influenced by the quality of fit versus the Occam penalty.

\begin{table*}
\centering
 \begin{tabular}{c | c c | c c } 
 \hline
 Model & $ {\displaystyle \ln {\cal{B}}}$ &$\Delta {\displaystyle \ln \mathcal{L}_\mathrm{max}}$ & $\chi_1, \chi_2$ identical? & binaries with $z<z^\text{min}$ \\
 \hline\hline
 \nonid & 0.00 & 0.00 & no & none \\
 \exdefault & $-0.06$ & $-0.39$ & yes & none \\
 \isp & $-0.70$ & $-0.47$ & yes & dynamical-like \\
 \nonid \isp & $-1.37$ & $-0.51$ & no & dynamical-like \\\hline
 \nonid with $\lambda_0=0$ & $-0.53$ & 1.04 & no & none \\
\exdefault with $\lambda_0=0$ & $-0.05$ & $-0.54$ & yes & none \\
 \exdefault with $z^\text{min}=-1$ & $-1.63$ & $-1.08$ & yes & none \\\hline
 \default & $-2.71$ & $-1.84$ & yes & yes \\
 \hline
 \end{tabular}
 \caption{Model selection results for the model variants summarized in Table~\ref{tab:variants} for GWTC-3 \textit{excluding GW191109}.}
 \label{tab:BF2}
\end{table*}

The preferred model variant with the highest Bayesian evidence is the \nonid variant, and so we measure Bayes factors with respect to this best-fit model.
The \default model is moderately disfavored with $\ln {\cal{B}} = -2.7$ (${\cal{B}}=0.067$)---a result consistent with Ref.~\cite{2021_better_spin}.
However, in contrast to \cite{2021_better_spin} (but consistent with \cite{no_evidence}), we find no strong preference for a sub-population of zero-spin binaries.
We attribute this difference to the technical issues summarized at the end of Section~\ref{sec:method}.
The \isp model is somewhat disfavored with $\ln {\cal{B}} = -0.70$ (${\cal{B}}=0.50$) suggesting a slight preference against models with a sub-population of dynamical mergers.
There is no significant preference for the other variants with $\ln {\cal{B}} > -0.06$ (${\cal{B}}>0.94$).
We observe no evidence that the primary spin distribution is different from the secondary spin distribution.
The two statistically significant conclusions from Table~\ref{tab:BF2} are that (1) the data prefer models with $z^\text{min} > -1$ over models with $z^\text{min}=-1$, and (2) the distribution of \ac{BBH} spin tilts is poorly described by the \default model.

In Table~\ref{tab:BF3}, we show model selection results obtained with GW191109.
The \default is still disfavored with $\ln {\cal{B}} = -1.33$ (${\cal{B}}=0.26$).
Since GW191109 exhibits support for $\chi_\text{eff}<0$, the model variant with $z^\text{min}=-1$ becomes the model with the highest Bayesian evidence.
Although models allowing for a negligible spin sub-population and flexible $z^\text{min}$ produce the highest maximum likelihood values, they incur an Occam penalty compared to models with $\lambda_0=0$ or $z^\text{min}=-1$, which means they do not produce the highest Bayes factors.
This illustrates that GW191109 by itself has an important affect on our results.
Further study is required in order to determine if parameter estimation results for this event are reliable given systematic uncertainties.

We note that in both Table~\ref{tab:BF3} and Table~\ref{tab:BF2}, the maximum likelihood for some nested model variants exceeds the maximum likelihood for the more general model variants. For example, in Table~\ref{tab:BF2}, the maximum likelihood for the \nonid with $\lambda_0 =0$ is larger than \nonid model which allows $\lambda_0$ at [0,1] interval. 
Since the former model variant is nested within the latter model variant, it should not produce a better fit.
We suspect this is due to undersampling when we fit more hyper-parameters in the \nonid model with the same set of posterior samples.
While we believe the Bayes factors and posterior distributions are reliable, the maximum likelihood values may be somewhat underestimated and should therefore be taken with a grain of salt.
Work is ongoing to achieve more thorough convergence.

\begin{table*}
\centering
 \begin{tabular}{c | c c | c c } 
 \hline
 Model & $ {\displaystyle \ln {\cal{B}}}$ &$\Delta {\displaystyle \ln \mathcal{L}_\mathrm{max}}$ & $\chi_1, \chi_2$ identical? & binaries with $z<z_\text{min}$ \\
 \hline\hline
 \exdefault & 0.00 & 0.00 & yes & none \\
 \isp & $-0.56$ & $-1.41$ & yes & dynamical-like \\
 \nonid & $-0.60$ & $-1.00$ & no & none \\
 \nonid \isp & $-0.64$ & $-0.22$ & no & dynamical-like \\\hline
\exdefault with $\lambda_0=0$ & 0.26 & $-1.46$ & yes & none \\
 \exdefault with $z_{min}=-1$ & 1.21 & $-2.35$ & yes & none \\\hline
 \default & $-1.33$ & $-2.14$ & yes & yes \\
 \hline
 \end{tabular}
 \caption{Model selection results for the model variants summarized in Table~\ref{tab:variants} for GWTC-3 \textit{including GW191109}.}
 \label{tab:BF3}
\end{table*}

Next, we carry out a comparison between the \exdefault model and $\chi_{\rm{eff}}$ \gaussian model in Refs.~\cite{no_evidence,gwtc2_pop}. 
Note that in the $\chi_\text{eff}$ \gaussian model variant, we only fit $\chi_\text{eff}$, but not the effective precession parameter $\chi_p$.
Thus, we adopt the same as priors used in parameter estimation for individual event.
Using data from GWTC-2, these two models were shown to produce qualitatively similar reconstructed distributions for $\chi_\text{eff}$ when no sub-population of zero-spin binaries is present ($\lambda_0=0$) \citep{gwtc2_pop}.
However, until now, it was not possible to compare the models directly because they were implemented with different analysis codes, and so we did not have Bayesian evidence values for both models.
For technical reasons, we include data only from \ac{GWTC-2}.
We find the \exdefault model is favored over the \gaussian model with $\ln{\cal{B}} = 7$ (${\cal{B}}=1100$) and $\Delta \ln \mathcal{L}_\text{max} \sim 5$. 
This suggests that \exdefault model provides significantly better fit than the \gaussian model.
Part of this result is likely driven by the $\chi_\text{eff}<0$ tail, which appears to contribute to the relatively poor fit of the \gaussian model.
Since we do not really fit $\chi_p$ in the \gaussian model, the \exdefault model may better fit the effective precession spin parameter $\chi_p$ as well (see Fig.~\ref{fig:chi_p}).


\subsection{Posterior distributions}
\begin{figure*}[htbp!]
    \centering
    \begin{subfigure}
        \centering
        \includegraphics[width=0.49\linewidth]{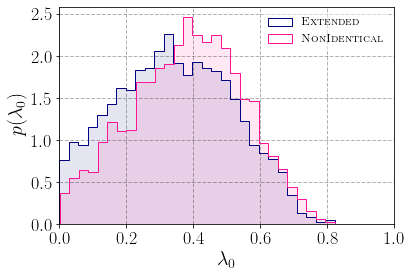}
    \end{subfigure}
    \hfill
    \begin{subfigure}
         \centering
         \includegraphics[width=0.49\linewidth]{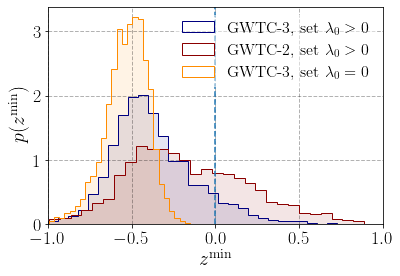}
    \end{subfigure}
    \caption{The posterior distributions for key population parameters. (We exclude GW191109.)
    Left is $\lambda_0$, the fraction of binaries with negligible black-hole spins.
    In this panel, different colors correspond to different model variants. We show only two traces here since posteriors for $\lambda_0$ of \isp model and \nonid \isp model are very similar to the traces from the \exdefault model and \nonid model respectively.
    We do not show the $\lambda_0=0$ posterior for GWTC-2 since it is similar to the posterior for $\lambda_0=0$ GWTC-3, just a bit broader.
    Both models show only a weak preference for $\lambda_0>0$.
    Right is \exdefault-model posterior for $z^\text{min}$, which controls the maximum spin misalignment angles.
    In this panel, the colors denote the dataset (GWTC-2 versus GWTC-3) and whether or not we assume a sub-population of \ac{BBH} mergers with zero spin. 
    We see that the support for $z_\text{min}<0$ depends strongly on the assumption that there is no sub-population with zero spin ($\lambda_0=0$).
    However, if we allow for non-spinning binaries, there is still modest evidence for anti-aligned binaries.
    }
    \label{fig:lam_chi_overplot}
\end{figure*}

A full corner plot for our best-fit model (\nonid) is provided in the Appendix (see Fig.~\ref{fig:corner_overplot})\footnote{Supplementary material including hyper-posterior samples for all model variants and result plots are available here: \url{https://github.com/HuiTong5/GWTC-3_pop_spin}}.
Of particular interest is the $\lambda_0$ hyper-parameter, which measures the fraction of \ac{BBH} mergers with non-spinning black holes.
In the left panel of Fig.~\ref{fig:lam_chi_overplot}, we plot the posterior for $\lambda_0$ for two model variants.
While every variant prefers $\lambda_0>0$, the statistical preference is weak when we take into account the Occam penalty for the introduction of the $\lambda_0$ parameter.
This supports previous conclusions that there is currently no evidence for or against a sub-population of binaries with negligible black-hole spin.
We strongly rule out $\lambda_0=1$, indicating that at least some \ac{BBH} systems contain spinning black holes, consistent with previous results \citep{Miller2020}.
Our credible 90\% interval $\lambda_0 = 0.39 ^{+ 0.20} _{- 0.24}$ (for the best-fit \nonid model) is now in broad agreement with \cite{no_evidence}, which gave an upper limit of $\lambda_0 \lesssim 65\%$.

Another parameter of interest is $z^\text{min}$, which affects the shape of the black hole spin tilt distribution.
In the right panel of Fig.~\ref{fig:lam_chi_overplot}, we plot the (\exdefault model variant) posterior distribution for $z^\text{min}$.
The result for GWTC-3 are shown in blue while the results for GWTC-2 are shown in red.
In orange, we show how the GWTC-3 result changes if we set $\lambda_0=0$ so that there is no sub-population of zero-spin \ac{BBH} mergers.
If we set $\lambda_0=0$, the data strongly favors $z^\text{min}<0$, which implies the existence of \ac{BBH} mergers with ``anti-aligned'' spin vectors (within the framework of these model variants).
However, consistent with results from Refs.~\cite{2021_better_spin,Roulet_2021}, the red trace shows that there was only weak evidence for $z^\text{min}<0$ ($69\%$ credibility) in the GWTC-2 catalog when we allow for a sub-population of black holes with zero spin; there is no compelling evidence for \ac{BBH} with ``anti-aligned'' spin vectors.

Turning our attention to the blue GWTC-3 trace, we find modest evidence for $z^\text{min}<0$ ($91\%$ credibility) when we allow for a sub-population of non-spinning black holes.
Repeating the analysis with the event GW191109 (to investigate currently unsubstantiated concerns about data quality), we find $z^\text{min}<0$ with $95 \%$ credibility (see Fig.~\ref{fig:zmin_lam_191109}).
We conclude that there is modest support in GWTC-3 for the hypothesis that some binaries merge with $\chi_\text{eff}<0$. We expect additional observations are required to determine if the signature is physical or a statistical fluctuation / model misspecification. It is interesting to compare and contrast our results with those from Ref~\cite{no_evidence}. Both analyses find strong evidence of $z^\text{min}<0$ when no zero-spin sub-population is allowed. However, in contrast to our study, Ref~\cite{no_evidence} still reports confident support for$z^\text{min}<0$ even when including a zero-spin sub-population. We speculate that this difference may come from different implementations of Monte Carlo averages. In our work, we employ a separate set of zero-spin samples, while Ref~\citep{no_evidence} represents each event’s posterior using a Gaussian kernel density estimate (KDE).

\begin{figure}[htbp!]
    \centering
    \includegraphics[width=1\columnwidth]{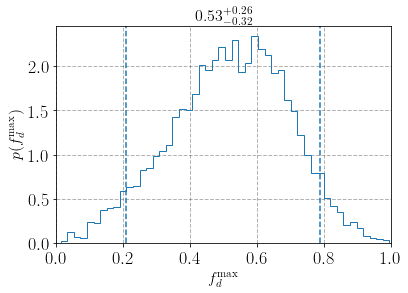}
    \caption{Posterior for the maximum fraction of dynamical mergers using \isp model variant. (GW191109 is excluded.)}
    \label{fig:max_dynamical}
\end{figure}

Thus, our results for $\lambda_0$ and $z_\text{min}$ are inconclusive.
The one astrophysical statement that we can make with some confidence is that at least some BBH systems seem to merge in the field with $\chi_\text{eff}>0$.
We ask: given our models, what is the largest possible fraction of mergers assembled dynamically?
Within the framework of the \isp variant, there are two sub-populations that have properties consistent with dynamical assembly: the sub-population of \ac{BBH} systems with no spin and the sub-population of \ac{BBH} systems with non-zero isotropic spin.
Of course, the zero-spin sub-population does not \textit{have to} be associated with dynamical assembly---this sub-population can also be associated with field binaries.
However, since so many caveats are possible, it is useful to frame things in terms of the maximum possible fraction of dynamically assembled binaries.
To this end, we calculate $f_d^\text{max}$--- the maximum fraction of dynamical mergers as determined by the \isp model variant:
\begin{equation}
    f_d^\text{max} = \lambda_0 + (1-\lambda_0)(1-\zeta) .
    \label{eq:max_dynamical}
\end{equation}
Here, $\lambda_0$ here corresponds to the fraction of mergers with no spin.
In Fig.~\ref{fig:max_dynamical} we plot the posterior distribution for the maximum fraction of dynamical mergers.
We find that $f_d^\text{max}\lesssim 89\%$ at $99\%$ credibility. This result is in broad agreement with the estimate of dynamical mergers in Ref~\cite{gwtc2_pop}, which finds the fraction of binaries arising from the dynamical channel to be $0.25 \leq f_\text{d} \leq 0.93$ at $90\%$ credibility, but more strongly suggesting that not all binaries merge dynamically (if we assume that dynamical assembly implies an isotropic distribution of spin vectors).
This is likely driven by the fact that the observed \ac{BBH} systems with clear signs of spin are all consistent with small spin tilt angles \cite{Roulet_2021}.
If we consider the possibility that all zero-spin BBH systems are formed in the field, then the \textit{minimum} \isp fraction of dynamical mergers is consistent with zero.

\subsection{Reconstructed distributions}
\begin{figure*}[htbp!]
    \begin{subfigure}
    \centering
    \includegraphics[width=0.49\linewidth]{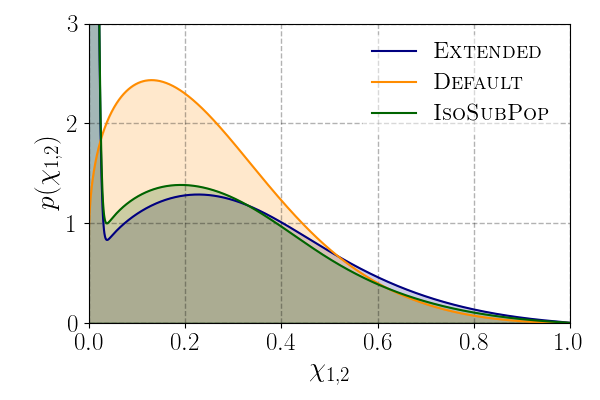}
    \end{subfigure}
    \hfill
    \begin{subfigure}
    \centering
    \includegraphics[width=0.49\linewidth]{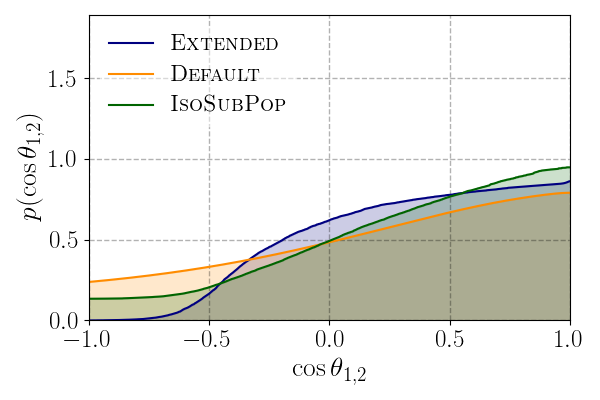}
    \end{subfigure}
    \caption{
    Population predictive distribution for different model variants; see Eq.~\ref{eq:ppd}. (We exclude GW191109 here.)
    Left shows the reconstructed distribution of dimensionless spin while right shows the reconstructed distribution of cosine tilt angle. Each color represents a different model variant from Table~\ref{tab:variants}.
    We included three typical models here. 
    The \nonid and \nonid \isp model variants are respectively similar to the \exdefault and \isp model variants. }
    \label{fig:ppd}
\end{figure*}

We now turn our attention to the reconstructed distributions for black hole spin implied by our fit.
The plots in this subsection exclude GW191109.
In Fig.~\ref{fig:ppd}, we plot the population predictive distribution (PPD) for dimensionless spin $\chi$ and cosine tilt angle $z$ given different model variants.
The PPD is calculated by marginalizing the prior over the posterior distribution of population parameters $\Lambda$:
\begin{equation}\label{eq:ppd}
p_\Lambda(\chi_{1,2}|d)=\int d\Lambda \,  p(\Lambda|d)\pi(\chi_{1,2}|\Lambda).  
\end{equation}
In the left-hand panel, \exdefault and \isp models show a $\chi=0$ spike, which to some extent account for the preference over \default model. The beta distribution includes only a small fraction of mergers with negligible $\chi \in (0, 0.01)$: $\lesssim 0.3\%$ (using maximum-likelihood hyper-parameter sample). So such spike is mostly contributed by the delta function sub-model.
The \default model does not appear to adequately fit this sharp feature.

In the right-hand panel of Fig.~\ref{fig:ppd}, we plot the PPD for cosine tilt angle $z$.
The result of the \default model clearly extends to very negative values, with $34\%$ of the distribution falling below $z<0$.
The \exdefault model are cut off at around $z\sim-0.6$.
In the \exdefault variant, 24\% of binaries merge with $z<0$; the number is 15\% for the \isp model.
Note that, unlike the \exdefault variant, the \isp model includes a realistic description of dynamical mergers with a truly isotropic orientation sub-population.
The model selection results suggest a slight preference for such $z$ cutoff when excluding GW191109. 
This is likely due to lack of observed events with unambiguously negative $\chi_\text{eff}$.

\begin{figure}
    \centering
    \includegraphics[width=1\columnwidth]{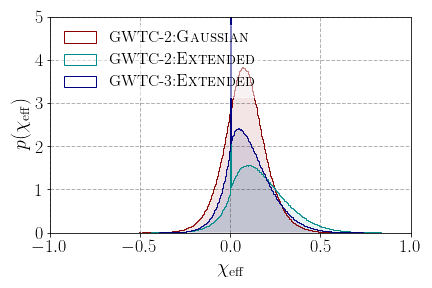}
    \caption{Population predictive reconstructed distribution for $\chi_\text{eff}$ of \gaussian and \exdefault model using GWTC-2/3 data. 
    The colors denote different combination of models and data. We exclude GW191109.}
    \label{fig:ppd_eff}
\end{figure}

In Fig.~\ref{fig:ppd_eff}, we show the PPD for the effective inspiral spin parameter $\chi_\text{eff}$.
We compare the results between the \gaussian model and the \exdefault model.
The different traces indicate which model is plotted and whether we use only GWTC-2 or GWTC-3.
These two models disagree most significantly in the region of $\chi_\text{eff} \lesssim 0.5$.
The peak at $\chi_\text{eff} \sim 0$ is consistent with the moderate support for non-vanishing $\lambda_0$ in the \exdefault model.
We also find an asymmetry in the $\chi_\text{eff}$ distribution using \exdefault model. 
These features are difficult to fit with the unimodal symmetric \gaussian model.
We include the reconstructed $\chi_\text{eff}$ for all the model variants in this work in the Appendix. 
The variant with the smallest PPD area with $\chi_\text{eff}<0$ is the \nonid variant with $\sim 9.8\%$.

\section{Discussions and conclusions}\label{sec:Discussion and conclusion}
In this work, we update the results from Ref.~\cite{2021_better_spin}, making corrections to that analysis, expanding the dataset to include GWTC-3, and considering an expanded set of model variants.
In agreement with Refs.~\cite{2021_better_spin, Roulet_2021}, we find that previous claims of anti-aligned black hole spin vectors \citep{gwtc2_pop} are model-dependent.
However, unlike Ref.~\cite{2021_better_spin}, we do not find clear evidence for a sub-population of zero-spin black holes; the current data are not sufficiently informative to determine if such a sub-population exists. 
This is in agreement with Ref.~\cite{no_evidence,Callister2021}.
We find modest support for BBH systems with $\chi_\text{eff}<0$.

Our estimate on the fraction of negligible-spin binaries are inconsistent with Ref.~\cite{mrr}, who conclude that only $\lesssim 1\%$ of \ac{BBH} systems merge with negligible spins for both the primary and secondary \ac{BH}. However, it is probably more fair to compare our estimate for the non-spinning fraction for the spinning sub-population, i.e., hyper-parameter $1-\lambda_0$ in our work, since models in Ref.~\cite{mrr} allow one spinning binaries. Ref.~\cite{mrr} reported $1-\lambda_0=0.77^{+0.16}_{-0.20}$, which is consistent with our result.
We endorse the idea of building models where no more than one black hole per binary has negligible spin, which is consistent with idea that some black hole progenitors are spun up through tides; see, e.g., Refs.~\cite{Broekgaarden2022,Qin2022,Mandel2020}.
It is possible that our results presented here are biased due to misspecification, and that we would find $\lambda_0 \approx 0$ if we allowed for sub-populations where at most one black hole spins.
Unfortunately, significant work is required to carry out further studies with ``single-spin'' models as considerable effort is required to generate primary and secondary single-spin posterior samples for each event.
Such dedicated samples may be necessary to avoid yet another form of bias arising from undersampling, which can become significant when trying to resolve sharp features in the population model.
Even so, the application of such ``single-spin'' models is a priority for future study.
A different spin parameterization method \cite{Biscoveanu_2021} may help improve the estimate of spin distribution in this scenario. Also, following \cite{Fuller_2019}, we model the sub-population of binary black holes with negligible spin using a delta function, which enforces zero spin. However, it may be that the true distribution is broader with support for small-but-non-zero spin as assumed in Refs.~\cite{no_evidence,mrr} which employ half-Gaussians with widths. Additional work is required in the future to study the nature of the purported sub-population of binary black holes with negligible spin. 

We find modest support for anti-aligned black hole spins with tilt angles $>90^\circ$---a result that requires subtle interpretation.
On the one hand, this result would seem to lend support to Ref.~\cite{gwtc2_pop}, which claimed some binaries merge with anti-aligned spin.
On the other hand, our results show that the conclusions drawn form GWTC-2 data analysis in Ref.~\cite{gwtc2_pop} are model-dependent because the evidence for anti-aligned spin is weak when we allow for a sub-population with negligible spin. 
Adding data from the latest LVK observing run (O3b), there is increased support for anti-aligned spin, even when we take into account the possibility of a sub-population with zero spin.
However, the statistical significance is modest. 
We do not find strong support for $\chi_\text{eff}<0$ as in Ref~\cite{no_evidence,2021arXiv211103606T}. In particular, Ref~\cite{no_evidence} confidently favors the existence of anti-aligned spin regardless of the presence of negligible spin sub-population.
Additional model misspecification may be lurking beneath the surface.
We therefore urge caution.

Putting everything together, we conclude that we are some ways away from determining the dominant channel for the \ac{BBH} mergers observed by the LVK.
There may or may not be a sub-population of \ac{BBH} systems with negligible spin.
There is modest evidence that some \ac{BBH} systems merge with anti-aligned spin, which could indicate dynamical assembly, though this signal could also be attributed to statistical fluctuations and/or model misspecification.
The one thing we think we can say confidently is that at least some LVK mergers must be assembled in the field: conservatively $\gtrsim 11\%$ (99\% credibility).

\begin{acknowledgments}
We thank Tom Callister, Sylvia Biscoveanu, and Jacob Lange for comments on an early draft of the manuscript.
We thank Colm Talbot, Ethan Payne, Katerina Chatziioannou, Jacob Golomb, Richard Udall, Derek Davis, Sophie Hourihane, and Paul Lasky for helpful discussions about the curious events GW200129 and GW191109.
This material is based upon work supported by NSF's LIGO Laboratory which is a major facility fully funded by the National Science Foundation. We are grateful for computational resources provided by the LIGO Laboratory and supported by National Science Foundation Grants PHY-0757058 and PHY-0823459.
The authors acknowledge support from the Australian Research Council (ARC) Centre of Excellence CE170100004.
The authors are grateful for computational resources provided by the LIGO Laboratory and supported by National Science Foundation Grants PHY-0757058 and PHY-0823459.
\end{acknowledgments}

\bibliography{refs}

\begin{thebibliography}{50}%
\makeatletter
\providecommand \@ifxundefined [1]{%
 \@ifx{#1\undefined}
}%
\providecommand \@ifnum [1]{%
 \ifnum #1\expandafter \@firstoftwo
 \else \expandafter \@secondoftwo
 \fi
}%
\providecommand \@ifx [1]{%
 \ifx #1\expandafter \@firstoftwo
 \else \expandafter \@secondoftwo
 \fi
}%
\providecommand \natexlab [1]{#1}%
\providecommand \enquote  [1]{``#1''}%
\providecommand \bibnamefont  [1]{#1}%
\providecommand \bibfnamefont [1]{#1}%
\providecommand \citenamefont [1]{#1}%
\providecommand \href@noop [0]{\@secondoftwo}%
\providecommand \href [0]{\begingroup \@sanitize@url \@href}%
\providecommand \@href[1]{\@@startlink{#1}\@@href}%
\providecommand \@@href[1]{\endgroup#1\@@endlink}%
\providecommand \@sanitize@url [0]{\catcode `\\12\catcode `\$12\catcode
  `\&12\catcode `\#12\catcode `\^12\catcode `\_12\catcode `\%12\relax}%
\providecommand \@@startlink[1]{}%
\providecommand \@@endlink[0]{}%
\providecommand \url  [0]{\begingroup\@sanitize@url \@url }%
\providecommand \@url [1]{\endgroup\@href {#1}{\urlprefix }}%
\providecommand \urlprefix  [0]{URL }%
\providecommand \Eprint [0]{\href }%
\providecommand \doibase [0]{http://dx.doi.org/}%
\providecommand \selectlanguage [0]{\@gobble}%
\providecommand \bibinfo  [0]{\@secondoftwo}%
\providecommand \bibfield  [0]{\@secondoftwo}%
\providecommand \translation [1]{[#1]}%
\providecommand \BibitemOpen [0]{}%
\providecommand \bibitemStop [0]{}%
\providecommand \bibitemNoStop [0]{.\EOS\space}%
\providecommand \EOS [0]{\spacefactor3000\relax}%
\providecommand \BibitemShut  [1]{\csname bibitem#1\endcsname}%
\let\auto@bib@innerbib\@empty
\bibitem [{\citenamefont {Abbott}\ \emph {et~al.}(2016)\citenamefont {Abbott},
  \citenamefont {Abbott}, \citenamefont {Abbott}, \citenamefont {Abernathy},
  \citenamefont {Acernese},\ and\ \citenamefont
  {et~al.}}]{PhysRevLett.116.061102}%
  \BibitemOpen
  \bibfield  {author} {\bibinfo {author} {\bibfnamefont {B.~P.}\ \bibnamefont
  {Abbott}}, \bibinfo {author} {\bibfnamefont {R.}~\bibnamefont {Abbott}},
  \bibinfo {author} {\bibfnamefont {T.~D.}\ \bibnamefont {Abbott}}, \bibinfo
  {author} {\bibfnamefont {M.~R.}\ \bibnamefont {Abernathy}}, \bibinfo {author}
  {\bibfnamefont {F.}~\bibnamefont {Acernese}}, \ and\ \bibinfo {author}
  {\bibnamefont {et~al.}} (\bibinfo {collaboration} {LIGO Scientific
  Collaboration and Virgo Collaboration}),\ }\bibfield  {title} {\enquote
  {\bibinfo {title} {Observation of gravitational waves from a binary black
  hole merger},}\ }\href {\doibase 10.1103/PhysRevLett.116.061102} {\bibfield
  {journal} {\bibinfo  {journal} {Phys. Rev. Lett.}\ }\textbf {\bibinfo
  {volume} {116}},\ \bibinfo {pages} {061102} (\bibinfo {year}
  {2016})}\BibitemShut {NoStop}%
\bibitem [{\citenamefont {{Abbott}}\ \emph {et~al.}(2021)\citenamefont
  {{Abbott}}, \citenamefont {{Abbott}}, \citenamefont {{Acernese}},
  \citenamefont {{Ackley}}, \citenamefont {{Adams}}, \citenamefont
  {{Adhikari}}, \citenamefont {{Adhikari}},\ and\ \citenamefont
  {et~al.}}]{2021arXiv211103606T}%
  \BibitemOpen
  \bibfield  {author} {\bibinfo {author} {\bibfnamefont {R.}~\bibnamefont
  {{Abbott}}}, \bibinfo {author} {\bibfnamefont {T.~D.}\ \bibnamefont
  {{Abbott}}}, \bibinfo {author} {\bibfnamefont {F.}~\bibnamefont
  {{Acernese}}}, \bibinfo {author} {\bibfnamefont {K.}~\bibnamefont
  {{Ackley}}}, \bibinfo {author} {\bibfnamefont {C.}~\bibnamefont {{Adams}}},
  \bibinfo {author} {\bibfnamefont {N.}~\bibnamefont {{Adhikari}}}, \bibinfo
  {author} {\bibfnamefont {R.~X.}\ \bibnamefont {{Adhikari}}}, \ and\ \bibinfo
  {author} {\bibnamefont {et~al.}},\ }\bibfield  {title} {\enquote {\bibinfo
  {title} {{GWTC-3: Compact Binary Coalescences Observed by LIGO and Virgo
  During the Second Part of the Third Observing Run}},}\ }\href@noop {}
  {\bibfield  {journal} {\bibinfo  {journal} {arXiv e-prints}\ ,\ \bibinfo
  {eid} {arXiv:2111.03606}} (\bibinfo {year} {2021})},\ \Eprint
  {http://arxiv.org/abs/2111.03606} {arXiv:2111.03606 [gr-qc]} \BibitemShut
  {NoStop}%
\bibitem [{\citenamefont {Nitz}\ \emph {et~al.}(2021)\citenamefont {Nitz},
  \citenamefont {Capano}, \citenamefont {Kumar}, \citenamefont {Wang},
  \citenamefont {Kastha}, \citenamefont {Schäfer}, \citenamefont {Dhurkunde},\
  and\ \citenamefont {Cabero}}]{Nitz_2021}%
  \BibitemOpen
  \bibfield  {author} {\bibinfo {author} {\bibfnamefont {Alexander~H.}\
  \bibnamefont {Nitz}}, \bibinfo {author} {\bibfnamefont {Collin~D.}\
  \bibnamefont {Capano}}, \bibinfo {author} {\bibfnamefont {Sumit}\
  \bibnamefont {Kumar}}, \bibinfo {author} {\bibfnamefont {Yi-Fan}\
  \bibnamefont {Wang}}, \bibinfo {author} {\bibfnamefont {Shilpa}\ \bibnamefont
  {Kastha}}, \bibinfo {author} {\bibfnamefont {Marlin}\ \bibnamefont
  {Schäfer}}, \bibinfo {author} {\bibfnamefont {Rahul}\ \bibnamefont
  {Dhurkunde}}, \ and\ \bibinfo {author} {\bibfnamefont {Miriam}\ \bibnamefont
  {Cabero}},\ }\bibfield  {title} {\enquote {\bibinfo {title} {3-{OGC}: Catalog
  of gravitational waves from compact-binary mergers},}\ }\href {\doibase
  10.3847/1538-4357/ac1c03} {\bibfield  {journal} {\bibinfo  {journal} {The
  Astrophysical Journal}\ }\textbf {\bibinfo {volume} {922}},\ \bibinfo {pages}
  {76} (\bibinfo {year} {2021})}\BibitemShut {NoStop}%
\bibitem [{\citenamefont {Olsen}\ \emph {et~al.}(2022)\citenamefont {Olsen},
  \citenamefont {Venumadhav}, \citenamefont {Mushkin}, \citenamefont {Roulet},
  \citenamefont {Zackay},\ and\ \citenamefont {Zaldarriaga}}]{Olsen_2022}%
  \BibitemOpen
  \bibfield  {author} {\bibinfo {author} {\bibfnamefont {Seth}\ \bibnamefont
  {Olsen}}, \bibinfo {author} {\bibfnamefont {Tejaswi}\ \bibnamefont
  {Venumadhav}}, \bibinfo {author} {\bibfnamefont {Jonathan}\ \bibnamefont
  {Mushkin}}, \bibinfo {author} {\bibfnamefont {Javier}\ \bibnamefont
  {Roulet}}, \bibinfo {author} {\bibfnamefont {Barak}\ \bibnamefont {Zackay}},
  \ and\ \bibinfo {author} {\bibfnamefont {Matias}\ \bibnamefont
  {Zaldarriaga}},\ }\href {\doibase 10.48550/ARXIV.2201.02252} {\enquote
  {\bibinfo {title} {New binary black hole mergers in the ligo--virgo o3a
  data},}\ } (\bibinfo {year} {2022})\BibitemShut {NoStop}%
\bibitem [{\citenamefont {Mapelli}(2020)}]{Mapelli_2020}%
  \BibitemOpen
  \bibfield  {author} {\bibinfo {author} {\bibfnamefont {Michela}\ \bibnamefont
  {Mapelli}},\ }\bibfield  {title} {\enquote {\bibinfo {title} {Binary black
  hole mergers: formation and populations},}\ }\href {\doibase
  10.3389/fspas.2020.00038} {\bibfield  {journal} {\bibinfo  {journal}
  {Frontiers in Astronomy and Space Sciences}\ }\textbf {\bibinfo {volume} {7}}
  (\bibinfo {year} {2020}),\ 10.3389/fspas.2020.00038}\BibitemShut {NoStop}%
\bibitem [{\citenamefont {Farr}\ \emph {et~al.}(2011)\citenamefont {Farr},
  \citenamefont {Kremer}, \citenamefont {Lyutikov},\ and\ \citenamefont
  {Kalogera}}]{Farr_2011}%
  \BibitemOpen
  \bibfield  {author} {\bibinfo {author} {\bibfnamefont {Will~M.}\ \bibnamefont
  {Farr}}, \bibinfo {author} {\bibfnamefont {Kyle}\ \bibnamefont {Kremer}},
  \bibinfo {author} {\bibfnamefont {Maxim}\ \bibnamefont {Lyutikov}}, \ and\
  \bibinfo {author} {\bibfnamefont {Vassiliki}\ \bibnamefont {Kalogera}},\
  }\bibfield  {title} {\enquote {\bibinfo {title} {Spin tilts in the double
  pulsar reveal supernova spin angular momentum production},}\ }\href {\doibase
  10.1088/0004-637x/742/2/81} {\bibfield  {journal} {\bibinfo  {journal}
  {Astrophys. J.}\ }\textbf {\bibinfo {volume} {742}},\ \bibinfo {pages} {81}
  (\bibinfo {year} {2011})}\BibitemShut {NoStop}%
\bibitem [{\citenamefont {Rodriguez}\ \emph {et~al.}(2016)\citenamefont
  {Rodriguez}, \citenamefont {Zevin}, \citenamefont {Pankow}, \citenamefont
  {Kalogera},\ and\ \citenamefont {Rasio}}]{Rodriguez_2016}%
  \BibitemOpen
  \bibfield  {author} {\bibinfo {author} {\bibfnamefont {Carl~L.}\ \bibnamefont
  {Rodriguez}}, \bibinfo {author} {\bibfnamefont {Michael}\ \bibnamefont
  {Zevin}}, \bibinfo {author} {\bibfnamefont {Chris}\ \bibnamefont {Pankow}},
  \bibinfo {author} {\bibfnamefont {Vasilliki}\ \bibnamefont {Kalogera}}, \
  and\ \bibinfo {author} {\bibfnamefont {Frederic~A.}\ \bibnamefont {Rasio}},\
  }\bibfield  {title} {\enquote {\bibinfo {title} {Illuminating black hole
  binary formation channels with spin in advanced ligo},}\ }\href {\doibase
  10.3847/2041-8205/832/1/l2} {\ \textbf {\bibinfo {volume} {832}},\ \bibinfo
  {pages} {L2} (\bibinfo {year} {2016})}\BibitemShut {NoStop}%
\bibitem [{\citenamefont {Vitale}\ \emph {et~al.}(2017)\citenamefont {Vitale},
  \citenamefont {Lynch}, \citenamefont {Sturani},\ and\ \citenamefont
  {Graff}}]{Vitale_2017}%
  \BibitemOpen
  \bibfield  {author} {\bibinfo {author} {\bibfnamefont {Salvatore}\
  \bibnamefont {Vitale}}, \bibinfo {author} {\bibfnamefont {Ryan}\ \bibnamefont
  {Lynch}}, \bibinfo {author} {\bibfnamefont {Riccardo}\ \bibnamefont
  {Sturani}}, \ and\ \bibinfo {author} {\bibfnamefont {Philip}\ \bibnamefont
  {Graff}},\ }\bibfield  {title} {\enquote {\bibinfo {title} {Use of
  gravitational waves to probe the formation channels of compact binaries},}\
  }\href {\doibase 10.1088/1361-6382/aa552e} {\bibfield  {journal} {\bibinfo
  {journal} {Classical and Quantum Gravity}\ }\textbf {\bibinfo {volume}
  {34}},\ \bibinfo {pages} {03LT01} (\bibinfo {year} {2017})}\BibitemShut
  {NoStop}%
\bibitem [{\citenamefont {Stevenson}\ \emph {et~al.}(2017)\citenamefont
  {Stevenson}, \citenamefont {Berry},\ and\ \citenamefont
  {Mandel}}]{Stevenson_2017}%
  \BibitemOpen
  \bibfield  {author} {\bibinfo {author} {\bibfnamefont {Simon}\ \bibnamefont
  {Stevenson}}, \bibinfo {author} {\bibfnamefont {Christopher P.~L.}\
  \bibnamefont {Berry}}, \ and\ \bibinfo {author} {\bibfnamefont {Ilya}\
  \bibnamefont {Mandel}},\ }\bibfield  {title} {\enquote {\bibinfo {title}
  {Hierarchical analysis of gravitational-wave measurements of binary black
  hole spin{\textendash}orbit misalignments},}\ }\href {\doibase
  10.1093/mnras/stx1764} {\bibfield  {journal} {\bibinfo  {journal} {Monthly
  Notices of the Royal Astronomical Society}\ }\textbf {\bibinfo {volume}
  {471}},\ \bibinfo {pages} {2801--2811} (\bibinfo {year} {2017})}\BibitemShut
  {NoStop}%
\bibitem [{\citenamefont {Farr}\ \emph {et~al.}(2017)\citenamefont {Farr},
  \citenamefont {Stevenson}, \citenamefont {Miller}, \citenamefont {Mandel},
  \citenamefont {Farr},\ and\ \citenamefont {Vecchio}}]{Farr_2017}%
  \BibitemOpen
  \bibfield  {author} {\bibinfo {author} {\bibfnamefont {Will~M.}\ \bibnamefont
  {Farr}}, \bibinfo {author} {\bibfnamefont {Simon}\ \bibnamefont {Stevenson}},
  \bibinfo {author} {\bibfnamefont {M.~Coleman}\ \bibnamefont {Miller}},
  \bibinfo {author} {\bibfnamefont {Ilya}\ \bibnamefont {Mandel}}, \bibinfo
  {author} {\bibfnamefont {Ben}\ \bibnamefont {Farr}}, \ and\ \bibinfo {author}
  {\bibfnamefont {Alberto}\ \bibnamefont {Vecchio}},\ }\bibfield  {title}
  {\enquote {\bibinfo {title} {Distinguishing spin-aligned and isotropic black
  hole populations with gravitational waves},}\ }\href {\doibase
  10.1038/nature23453} {\bibfield  {journal} {\bibinfo  {journal} {Nature}\
  }\textbf {\bibinfo {volume} {548}},\ \bibinfo {pages} {426--429} (\bibinfo
  {year} {2017})}\BibitemShut {NoStop}%
\bibitem [{\citenamefont {Talbot}\ and\ \citenamefont
  {Thrane}(2017)}]{Talbot_2017}%
  \BibitemOpen
  \bibfield  {author} {\bibinfo {author} {\bibfnamefont {Colm}\ \bibnamefont
  {Talbot}}\ and\ \bibinfo {author} {\bibfnamefont {Eric}\ \bibnamefont
  {Thrane}},\ }\bibfield  {title} {\enquote {\bibinfo {title} {Determining the
  population properties of spinning black holes},}\ }\href {\doibase
  10.1103/physrevd.96.023012} {\bibfield  {journal} {\bibinfo  {journal}
  {Physical Review D}\ }\textbf {\bibinfo {volume} {96}} (\bibinfo {year}
  {2017}),\ 10.1103/physrevd.96.023012}\BibitemShut {NoStop}%
\bibitem [{\citenamefont {Gerosa}\ \emph {et~al.}(2018)\citenamefont {Gerosa},
  \citenamefont {Berti}, \citenamefont {O'Shaughnessy}, \citenamefont
  {Belczynski}, \citenamefont {Kesden}, \citenamefont {Wysocki},\ and\
  \citenamefont {Gladysz}}]{Gerosa_2018}%
  \BibitemOpen
  \bibfield  {author} {\bibinfo {author} {\bibfnamefont {Davide}\ \bibnamefont
  {Gerosa}}, \bibinfo {author} {\bibfnamefont {Emanuele}\ \bibnamefont
  {Berti}}, \bibinfo {author} {\bibfnamefont {Richard}\ \bibnamefont
  {O'Shaughnessy}}, \bibinfo {author} {\bibfnamefont {Krzysztof}\ \bibnamefont
  {Belczynski}}, \bibinfo {author} {\bibfnamefont {Michael}\ \bibnamefont
  {Kesden}}, \bibinfo {author} {\bibfnamefont {Daniel}\ \bibnamefont
  {Wysocki}}, \ and\ \bibinfo {author} {\bibfnamefont {Wojciech}\ \bibnamefont
  {Gladysz}},\ }\bibfield  {title} {\enquote {\bibinfo {title} {Spin
  orientations of merging black holes formed from the evolution of stellar
  binaries},}\ }\href {\doibase 10.1103/physrevd.98.084036} {\bibfield
  {journal} {\bibinfo  {journal} {Physical Review D}\ }\textbf {\bibinfo
  {volume} {98}} (\bibinfo {year} {2018}),\
  10.1103/physrevd.98.084036}\BibitemShut {NoStop}%
\bibitem [{\citenamefont {Franciolini}\ and\ \citenamefont
  {Pani}(2022)}]{Franciolini:2022iaa}%
  \BibitemOpen
  \bibfield  {author} {\bibinfo {author} {\bibfnamefont {Gabriele}\
  \bibnamefont {Franciolini}}\ and\ \bibinfo {author} {\bibfnamefont {Paolo}\
  \bibnamefont {Pani}},\ }\bibfield  {title} {\enquote {\bibinfo {title}
  {{Searching for mass-spin correlations in the population of
  gravitational-wave events: The GWTC-3 case study}},}\ }\href {\doibase
  10.1103/PhysRevD.105.123024} {\bibfield  {journal} {\bibinfo  {journal}
  {Phys. Rev. D}\ }\textbf {\bibinfo {volume} {105}},\ \bibinfo {pages}
  {123024} (\bibinfo {year} {2022})},\ \Eprint
  {http://arxiv.org/abs/2201.13098} {arXiv:2201.13098 [astro-ph.HE]}
  \BibitemShut {NoStop}%
\bibitem [{\citenamefont {Callister}\ \emph
  {et~al.}(2021{\natexlab{a}})\citenamefont {Callister}, \citenamefont
  {Haster}, \citenamefont {Ng}, \citenamefont {Vitale},\ and\ \citenamefont
  {Farr}}]{Callister_2021}%
  \BibitemOpen
  \bibfield  {author} {\bibinfo {author} {\bibfnamefont {Thomas~A.}\
  \bibnamefont {Callister}}, \bibinfo {author} {\bibfnamefont {Carl-Johan}\
  \bibnamefont {Haster}}, \bibinfo {author} {\bibfnamefont {Ken K.~Y.}\
  \bibnamefont {Ng}}, \bibinfo {author} {\bibfnamefont {Salvatore}\
  \bibnamefont {Vitale}}, \ and\ \bibinfo {author} {\bibfnamefont {Will~M.}\
  \bibnamefont {Farr}},\ }\bibfield  {title} {\enquote {\bibinfo {title} {Who
  ordered that? unequal-mass binary black hole mergers have larger effective
  spins},}\ }\href {\doibase 10.3847/2041-8213/ac2ccc} {\bibfield  {journal}
  {\bibinfo  {journal} {The Astrophysical Journal Letters}\ }\textbf {\bibinfo
  {volume} {922}},\ \bibinfo {pages} {L5} (\bibinfo {year}
  {2021}{\natexlab{a}})}\BibitemShut {NoStop}%
\bibitem [{\citenamefont {Adamcewicz}\ and\ \citenamefont
  {Thrane}(2022)}]{Do_unequal}%
  \BibitemOpen
  \bibfield  {author} {\bibinfo {author} {\bibfnamefont {Christian}\
  \bibnamefont {Adamcewicz}}\ and\ \bibinfo {author} {\bibfnamefont {Eric}\
  \bibnamefont {Thrane}},\ }\href {\doibase 10.48550/ARXIV.2208.03405}
  {\enquote {\bibinfo {title} {Do unequal-mass binary black hole systems have
  larger $\chi_\text{eff}$? probing correlations with copulas in
  gravitational-wave astronomy},}\ } (\bibinfo {year} {2022})\BibitemShut
  {NoStop}%
\bibitem [{\citenamefont {Abbott}\ \emph
  {et~al.}(2021{\natexlab{a}})\citenamefont {Abbott}, \citenamefont {Abbott},
  \citenamefont {Abraham}, \citenamefont {Acernese},\ and\ \citenamefont
  {et~al.}}]{gwtc2_pop}%
  \BibitemOpen
  \bibfield  {author} {\bibinfo {author} {\bibfnamefont {R.}~\bibnamefont
  {Abbott}}, \bibinfo {author} {\bibfnamefont {T.~D.}\ \bibnamefont {Abbott}},
  \bibinfo {author} {\bibfnamefont {S.}~\bibnamefont {Abraham}}, \bibinfo
  {author} {\bibfnamefont {F.}~\bibnamefont {Acernese}}, \ and\ \bibinfo
  {author} {\bibnamefont {et~al.}},\ }\bibfield  {title} {\enquote {\bibinfo
  {title} {Population properties of compact objects from the second
  {LIGO}{\textendash}virgo gravitational-wave transient catalog},}\ }\href
  {\doibase 10.3847/2041-8213/abe949} {\bibfield  {journal} {\bibinfo
  {journal} {The Astrophysical Journal Letters}\ }\textbf {\bibinfo {volume}
  {913}},\ \bibinfo {pages} {L7} (\bibinfo {year}
  {2021}{\natexlab{a}})}\BibitemShut {NoStop}%
\bibitem [{\citenamefont {Galaudage}\ \emph {et~al.}(2021)\citenamefont
  {Galaudage}, \citenamefont {Talbot}, \citenamefont {Nagar}, \citenamefont
  {Jain}, \citenamefont {Thrane},\ and\ \citenamefont
  {Mandel}}]{2021_better_spin}%
  \BibitemOpen
  \bibfield  {author} {\bibinfo {author} {\bibfnamefont {Shanika}\ \bibnamefont
  {Galaudage}}, \bibinfo {author} {\bibfnamefont {Colm}\ \bibnamefont
  {Talbot}}, \bibinfo {author} {\bibfnamefont {Tushar}\ \bibnamefont {Nagar}},
  \bibinfo {author} {\bibfnamefont {Deepnika}\ \bibnamefont {Jain}}, \bibinfo
  {author} {\bibfnamefont {Eric}\ \bibnamefont {Thrane}}, \ and\ \bibinfo
  {author} {\bibfnamefont {Ilya}\ \bibnamefont {Mandel}},\ }\bibfield  {title}
  {\enquote {\bibinfo {title} {Building better spin models for merging binary
  black holes: Evidence for nonspinning and rapidly spinning nearly aligned
  subpopulations},}\ }\href {\doibase 10.3847/2041-8213/ac2f3c} {\bibfield
  {journal} {\bibinfo  {journal} {The Astrophysical Journal Letters}\ }\textbf
  {\bibinfo {volume} {921}},\ \bibinfo {pages} {L15} (\bibinfo {year}
  {2021})}\BibitemShut {NoStop}%
\bibitem [{\citenamefont {Abbott}\ \emph
  {et~al.}(2021{\natexlab{b}})\citenamefont {Abbott} \emph
  {et~al.}}]{LIGOScientific:2021psn}%
  \BibitemOpen
  \bibfield  {author} {\bibinfo {author} {\bibfnamefont {R.}~\bibnamefont
  {Abbott}} \emph {et~al.} (\bibinfo {collaboration} {LIGO Scientific, VIRGO,
  KAGRA}),\ }\bibfield  {title} {\enquote {\bibinfo {title} {{The population of
  merging compact binaries inferred using gravitational waves through
  GWTC-3}},}\ }\href@noop {} {\  (\bibinfo {year} {2021}{\natexlab{b}})},\
  \Eprint {http://arxiv.org/abs/2111.03634} {arXiv:2111.03634 [astro-ph.HE]}
  \BibitemShut {NoStop}%
\bibitem [{\citenamefont {Callister}\ \emph {et~al.}(2022)\citenamefont
  {Callister}, \citenamefont {Miller}, \citenamefont {Chatziioannou},\ and\
  \citenamefont {Farr}}]{no_evidence}%
  \BibitemOpen
  \bibfield  {author} {\bibinfo {author} {\bibfnamefont {Thomas~A.}\
  \bibnamefont {Callister}}, \bibinfo {author} {\bibfnamefont {Simona~J.}\
  \bibnamefont {Miller}}, \bibinfo {author} {\bibfnamefont {Katerina}\
  \bibnamefont {Chatziioannou}}, \ and\ \bibinfo {author} {\bibfnamefont
  {Will~M.}\ \bibnamefont {Farr}},\ }\href {\doibase 10.48550/ARXIV.2205.08574}
  {\enquote {\bibinfo {title} {No evidence that the majority of black holes in
  binaries have zero spin},}\ } (\bibinfo {year} {2022})\BibitemShut {NoStop}%
\bibitem [{\citenamefont {Mould}\ \emph {et~al.}(2022)\citenamefont {Mould},
  \citenamefont {Gerosa}, \citenamefont {Broekgaarden},\ and\ \citenamefont
  {Steinle}}]{mrr}%
  \BibitemOpen
  \bibfield  {author} {\bibinfo {author} {\bibfnamefont {Matthew}\ \bibnamefont
  {Mould}}, \bibinfo {author} {\bibfnamefont {Davide}\ \bibnamefont {Gerosa}},
  \bibinfo {author} {\bibfnamefont {Floor~S.}\ \bibnamefont {Broekgaarden}}, \
  and\ \bibinfo {author} {\bibfnamefont {Nathan}\ \bibnamefont {Steinle}},\
  }\href {\doibase 10.48550/ARXIV.2205.12329} {\enquote {\bibinfo {title}
  {Which black hole formed first? mass-ratio reversal in massive binary stars
  from gravitational-wave data},}\ } (\bibinfo {year} {2022})\BibitemShut
  {NoStop}%
\bibitem [{\citenamefont {Biscoveanu}\ \emph {et~al.}(2022)\citenamefont
  {Biscoveanu}, \citenamefont {Callister}, \citenamefont {Haster},
  \citenamefont {Ng}, \citenamefont {Vitale},\ and\ \citenamefont
  {Farr}}]{Biscoveanu_2022}%
  \BibitemOpen
  \bibfield  {author} {\bibinfo {author} {\bibfnamefont {Sylvia}\ \bibnamefont
  {Biscoveanu}}, \bibinfo {author} {\bibfnamefont {Thomas~A.}\ \bibnamefont
  {Callister}}, \bibinfo {author} {\bibfnamefont {Carl-Johan}\ \bibnamefont
  {Haster}}, \bibinfo {author} {\bibfnamefont {Ken K.~Y.}\ \bibnamefont {Ng}},
  \bibinfo {author} {\bibfnamefont {Salvatore}\ \bibnamefont {Vitale}}, \ and\
  \bibinfo {author} {\bibfnamefont {Will~M.}\ \bibnamefont {Farr}},\ }\bibfield
   {title} {\enquote {\bibinfo {title} {The binary black hole spin distribution
  likely broadens with redshift},}\ }\href {\doibase 10.3847/2041-8213/ac71a8}
  {\bibfield  {journal} {\bibinfo  {journal} {The Astrophysical Journal
  Letters}\ }\textbf {\bibinfo {volume} {932}},\ \bibinfo {pages} {L19}
  (\bibinfo {year} {2022})}\BibitemShut {NoStop}%
\bibitem [{\citenamefont {Fishbach}\ \emph {et~al.}(2022)\citenamefont
  {Fishbach}, \citenamefont {Kimball},\ and\ \citenamefont
  {Kalogera}}]{Fishbach_2022}%
  \BibitemOpen
  \bibfield  {author} {\bibinfo {author} {\bibfnamefont {Maya}\ \bibnamefont
  {Fishbach}}, \bibinfo {author} {\bibfnamefont {Chase}\ \bibnamefont
  {Kimball}}, \ and\ \bibinfo {author} {\bibfnamefont {Vicky}\ \bibnamefont
  {Kalogera}},\ }\bibfield  {title} {\enquote {\bibinfo {title} {Limits on
  hierarchical black hole mergers from the most negative $\chi_\text{eff}$
  systems},}\ }\href {\doibase 10.3847/2041-8213/ac86c4} {\bibfield  {journal}
  {\bibinfo  {journal} {The Astrophysical Journal Letters}\ }\textbf {\bibinfo
  {volume} {935}},\ \bibinfo {pages} {L26} (\bibinfo {year}
  {2022})}\BibitemShut {NoStop}%
\bibitem [{\citenamefont {{LIGO Scientific Collaboration}}\ \emph
  {et~al.}(2015)\citenamefont {{LIGO Scientific Collaboration}}, \citenamefont
  {{Aasi}}, \citenamefont {{Abbott}}, \citenamefont {{Abbott}}, \citenamefont
  {{Abbott}}, \citenamefont {{Abernathy}}, \citenamefont {{Ackley}} \emph
  {et~al.}}]{AdLIGO}%
  \BibitemOpen
  \bibfield  {author} {\bibinfo {author} {\bibnamefont {{LIGO Scientific
  Collaboration}}}, \bibinfo {author} {\bibfnamefont {J.}~\bibnamefont
  {{Aasi}}}, \bibinfo {author} {\bibfnamefont {B.~P.}\ \bibnamefont
  {{Abbott}}}, \bibinfo {author} {\bibfnamefont {R.}~\bibnamefont {{Abbott}}},
  \bibinfo {author} {\bibfnamefont {T.}~\bibnamefont {{Abbott}}}, \bibinfo
  {author} {\bibfnamefont {M.~R.}\ \bibnamefont {{Abernathy}}}, \bibinfo
  {author} {\bibfnamefont {K.}~\bibnamefont {{Ackley}}},  \emph {et~al.},\
  }\bibfield  {title} {\enquote {\bibinfo {title} {{Advanced LIGO}},}\ }\href
  {\doibase 10.1088/0264-9381/32/7/074001} {\bibfield  {journal} {\bibinfo
  {journal} {Class. Quantum Grav.}\ }\textbf {\bibinfo {volume} {32}},\
  \bibinfo {eid} {074001} (\bibinfo {year} {2015})}\BibitemShut {NoStop}%
\bibitem [{\citenamefont {{Acernese}}\ \emph {et~al.}(2015)\citenamefont
  {{Acernese}} \emph {et~al.}}]{Virgo}%
  \BibitemOpen
  \bibfield  {author} {\bibinfo {author} {\bibfnamefont {F.}~\bibnamefont
  {{Acernese}}} \emph {et~al.},\ }\bibfield  {title} {\enquote {\bibinfo
  {title} {{Advanced Virgo: a second-generation interferometric gravitational
  wave detector}},}\ }\href@noop {} {\bibfield  {journal} {\bibinfo  {journal}
  {Class. Quantum Grav.}\ }\textbf {\bibinfo {volume} {32}},\ \bibinfo {eid}
  {024001} (\bibinfo {year} {2015})}\BibitemShut {NoStop}%
\bibitem [{\citenamefont {Damour}(2001)}]{Damour_2001}%
  \BibitemOpen
  \bibfield  {author} {\bibinfo {author} {\bibfnamefont {Thibault}\
  \bibnamefont {Damour}},\ }\bibfield  {title} {\enquote {\bibinfo {title}
  {Coalescence of two spinning black holes: An effective one-body approach},}\
  }\href {\doibase 10.1103/physrevd.64.124013} {\bibfield  {journal} {\bibinfo
  {journal} {Physical Review D}\ }\textbf {\bibinfo {volume} {64}} (\bibinfo
  {year} {2001}),\ 10.1103/physrevd.64.124013}\BibitemShut {NoStop}%
\bibitem [{\citenamefont {Ajith}\ \emph {et~al.}(2011)\citenamefont {Ajith},
  \citenamefont {Hannam}, \citenamefont {Husa}, \citenamefont {Chen},
  \citenamefont {Brügmann}, \citenamefont {Dorband}, \citenamefont {Müller},
  \citenamefont {Ohme}, \citenamefont {Pollney}, \citenamefont {Reisswig},
  \citenamefont {Santamar{\'{\i} }a},\ and\ \citenamefont
  {Seiler}}]{Ajith_2011}%
  \BibitemOpen
  \bibfield  {author} {\bibinfo {author} {\bibfnamefont {P.}~\bibnamefont
  {Ajith}}, \bibinfo {author} {\bibfnamefont {M.}~\bibnamefont {Hannam}},
  \bibinfo {author} {\bibfnamefont {S.}~\bibnamefont {Husa}}, \bibinfo {author}
  {\bibfnamefont {Y.}~\bibnamefont {Chen}}, \bibinfo {author} {\bibfnamefont
  {B.}~\bibnamefont {Brügmann}}, \bibinfo {author} {\bibfnamefont
  {N.}~\bibnamefont {Dorband}}, \bibinfo {author} {\bibfnamefont
  {D.}~\bibnamefont {Müller}}, \bibinfo {author} {\bibfnamefont
  {F.}~\bibnamefont {Ohme}}, \bibinfo {author} {\bibfnamefont {D.}~\bibnamefont
  {Pollney}}, \bibinfo {author} {\bibfnamefont {C.}~\bibnamefont {Reisswig}},
  \bibinfo {author} {\bibfnamefont {L.}~\bibnamefont {Santamar{\'{\i} }a}}, \
  and\ \bibinfo {author} {\bibfnamefont {J.}~\bibnamefont {Seiler}},\
  }\bibfield  {title} {\enquote {\bibinfo {title} {Inspiral-merger-ringdown
  waveforms for black-hole binaries with nonprecessing spins},}\ }\href
  {\doibase 10.1103/physrevlett.106.241101} {\bibfield  {journal} {\bibinfo
  {journal} {Physical Review Letters}\ }\textbf {\bibinfo {volume} {106}}
  (\bibinfo {year} {2011}),\ 10.1103/physrevlett.106.241101}\BibitemShut
  {NoStop}%
\bibitem [{\citenamefont {Stevenson}(2022)}]{Stevenson_2022}%
  \BibitemOpen
  \bibfield  {author} {\bibinfo {author} {\bibfnamefont {S}~\bibnamefont
  {Stevenson}},\ }\bibfield  {title} {\enquote {\bibinfo {title} {Biases in
  estimates of black hole kicks from the spin distribution of binary black
  holes},}\ }\href@noop {} {\bibfield  {journal} {\bibinfo  {journal}
  {Astrophys. J. Lett.}\ }\textbf {\bibinfo {volume} {926}},\ \bibinfo {pages}
  {L32} (\bibinfo {year} {2022})}\BibitemShut {NoStop}%
\bibitem [{\citenamefont {Roulet}\ \emph {et~al.}(2021)\citenamefont {Roulet},
  \citenamefont {Chia}, \citenamefont {Olsen}, \citenamefont {Dai},
  \citenamefont {Venumadhav}, \citenamefont {Zackay},\ and\ \citenamefont
  {Zaldarriaga}}]{Roulet_2021}%
  \BibitemOpen
  \bibfield  {author} {\bibinfo {author} {\bibfnamefont {Javier}\ \bibnamefont
  {Roulet}}, \bibinfo {author} {\bibfnamefont {Horng~Sheng}\ \bibnamefont
  {Chia}}, \bibinfo {author} {\bibfnamefont {Seth}\ \bibnamefont {Olsen}},
  \bibinfo {author} {\bibfnamefont {Liang}\ \bibnamefont {Dai}}, \bibinfo
  {author} {\bibfnamefont {Tejaswi}\ \bibnamefont {Venumadhav}}, \bibinfo
  {author} {\bibfnamefont {Barak}\ \bibnamefont {Zackay}}, \ and\ \bibinfo
  {author} {\bibfnamefont {Matias}\ \bibnamefont {Zaldarriaga}},\ }\bibfield
  {title} {\enquote {\bibinfo {title} {Distribution of effective spins and
  masses of binary black holes from the {LIGO} and virgo o1{\textendash}o3a
  observing runs},}\ }\href@noop {} {\bibfield  {journal} {\bibinfo  {journal}
  {Phys. Rev. D}\ }\textbf {\bibinfo {volume} {104}},\ \bibinfo {pages}
  {083010} (\bibinfo {year} {2021})}\BibitemShut {NoStop}%
\bibitem [{\citenamefont {Romero-Shaw}\ \emph {et~al.}(2022)\citenamefont
  {Romero-Shaw}, \citenamefont {Thrane},\ and\ \citenamefont {Lasky}}]{wmf}%
  \BibitemOpen
  \bibfield  {author} {\bibinfo {author} {\bibfnamefont {I.~M.}\ \bibnamefont
  {Romero-Shaw}}, \bibinfo {author} {\bibfnamefont {Eric}\ \bibnamefont
  {Thrane}}, \ and\ \bibinfo {author} {\bibfnamefont {Paul~D.}\ \bibnamefont
  {Lasky}},\ }\bibfield  {title} {\enquote {\bibinfo {title} {When models fail:
  an introduction to posterior predictive checks and model misspecification in
  gravitational-wave astronomy},}\ }\href@noop {} {\bibfield  {journal}
  {\bibinfo  {journal} {Pub. Astron. Soc. Aust.}\ }\textbf {\bibinfo {volume}
  {39}},\ \bibinfo {pages} {E025} (\bibinfo {year} {2022})}\BibitemShut
  {NoStop}%
\bibitem [{\citenamefont {Payne}\ \emph {et~al.}(2022)\citenamefont {Payne},
  \citenamefont {Hourihane}, \citenamefont {Golomb}, \citenamefont {Udall},
  \citenamefont {Davis},\ and\ \citenamefont {Chatziioannou}}]{Payne2022}%
  \BibitemOpen
  \bibfield  {author} {\bibinfo {author} {\bibfnamefont {Ethan}\ \bibnamefont
  {Payne}}, \bibinfo {author} {\bibfnamefont {Sophie}\ \bibnamefont
  {Hourihane}}, \bibinfo {author} {\bibfnamefont {Jacob}\ \bibnamefont
  {Golomb}}, \bibinfo {author} {\bibfnamefont {Richard}\ \bibnamefont {Udall}},
  \bibinfo {author} {\bibfnamefont {Derek}\ \bibnamefont {Davis}}, \ and\
  \bibinfo {author} {\bibfnamefont {Katerina}\ \bibnamefont {Chatziioannou}},\
  }\bibfield  {title} {\enquote {\bibinfo {title} {The curious case of
  gw200129: interplay between spin-precession inference and data-quality
  issues},}\ }\href@noop {} {\  (\bibinfo {year} {2022})},\ \bibinfo {note}
  {arxiv/2206.11932}\BibitemShut {NoStop}%
\bibitem [{\citenamefont {Hannam}\ \emph {et~al.}(2022)\citenamefont {Hannam}
  \emph {et~al.}}]{Hannam2022}%
  \BibitemOpen
  \bibfield  {author} {\bibinfo {author} {\bibfnamefont {Mark}\ \bibnamefont
  {Hannam}} \emph {et~al.},\ }\bibfield  {title} {\enquote {\bibinfo {title}
  {Measurement of general-relativistic precession in a black-hole binary},}\
  }\href@noop {} {\  (\bibinfo {year} {2022})},\ \bibinfo {note}
  {arxiv/2112.11300}\BibitemShut {NoStop}%
\bibitem [{Note1()}]{Note1}%
  \BibitemOpen
  \bibinfo {note} {For convenience, we re-parameterize the Beta distribution in
  terms of the spin magnitude mean $\mu _\chi $ and standard deviation $\sigma
  _\chi $ in our population analyses realization.}\BibitemShut {Stop}%
\bibitem [{\citenamefont {Fuller}\ and\ \citenamefont
  {Ma}(2019)}]{Fuller_2019}%
  \BibitemOpen
  \bibfield  {author} {\bibinfo {author} {\bibfnamefont {Jim}\ \bibnamefont
  {Fuller}}\ and\ \bibinfo {author} {\bibfnamefont {Linhao}\ \bibnamefont
  {Ma}},\ }\bibfield  {title} {\enquote {\bibinfo {title} {Most black holes are
  born very slowly rotating},}\ }\href {\doibase 10.3847/2041-8213/ab339b}
  {\bibfield  {journal} {\bibinfo  {journal} {The Astrophysical Journal}\
  }\textbf {\bibinfo {volume} {881}},\ \bibinfo {pages} {L1} (\bibinfo {year}
  {2019})}\BibitemShut {NoStop}%
\bibitem [{Note2()}]{Note2}%
  \BibitemOpen
  \bibinfo {note} {We use the phrase ``quasi-isotropically'' because the
  \protect \textsc {Extended}\protect \xspace and the \protect \textsc
  {NonIdentical}\protect \xspace model variants truncate the spin tilt
  distribution for dynamical-like binaries, and so the spin tilt distribution
  is not actually isotropic.}\BibitemShut {Stop}%
\bibitem [{\citenamefont {Talbot}\ and\ \citenamefont
  {Thrane}(2018)}]{2018_colm}%
  \BibitemOpen
  \bibfield  {author} {\bibinfo {author} {\bibfnamefont {Colm}\ \bibnamefont
  {Talbot}}\ and\ \bibinfo {author} {\bibfnamefont {Eric}\ \bibnamefont
  {Thrane}},\ }\bibfield  {title} {\enquote {\bibinfo {title} {Measuring the
  binary black hole mass spectrum with an astrophysically motivated
  parameterization},}\ }\href {\doibase 10.3847/1538-4357/aab34c} {\bibfield
  {journal} {\bibinfo  {journal} {The Astrophysical Journal}\ }\textbf
  {\bibinfo {volume} {856}},\ \bibinfo {pages} {173} (\bibinfo {year}
  {2018})}\BibitemShut {NoStop}%
\bibitem [{\citenamefont {Fishbach}\ \emph {et~al.}(2018)\citenamefont
  {Fishbach}, \citenamefont {Holz},\ and\ \citenamefont {Farr}}]{2018_maya}%
  \BibitemOpen
  \bibfield  {author} {\bibinfo {author} {\bibfnamefont {Maya}\ \bibnamefont
  {Fishbach}}, \bibinfo {author} {\bibfnamefont {Daniel~E.}\ \bibnamefont
  {Holz}}, \ and\ \bibinfo {author} {\bibfnamefont {Will~M.}\ \bibnamefont
  {Farr}},\ }\bibfield  {title} {\enquote {\bibinfo {title} {Does the black
  hole merger rate evolve with redshift?}}\ }\href {\doibase
  10.3847/2041-8213/aad800} {\bibfield  {journal} {\bibinfo  {journal} {The
  Astrophysical Journal}\ }\textbf {\bibinfo {volume} {863}},\ \bibinfo {pages}
  {L41} (\bibinfo {year} {2018})}\BibitemShut {NoStop}%
\bibitem [{Note3()}]{Note3}%
  \BibitemOpen
  \bibinfo {note} {\protect \url
  {https://zenodo.org/record/5546663}}\BibitemShut {NoStop}%
\bibitem [{\citenamefont {{Talbot}}\ \emph {et~al.}(2019)\citenamefont
  {{Talbot}}, \citenamefont {{Smith}}, \citenamefont {{Thrane}},\ and\
  \citenamefont {{Poole}}}]{2019PhRvD.100d3030T}%
  \BibitemOpen
  \bibfield  {author} {\bibinfo {author} {\bibfnamefont {Colm}\ \bibnamefont
  {{Talbot}}}, \bibinfo {author} {\bibfnamefont {Rory}\ \bibnamefont
  {{Smith}}}, \bibinfo {author} {\bibfnamefont {Eric}\ \bibnamefont
  {{Thrane}}}, \ and\ \bibinfo {author} {\bibfnamefont {Gregory~B.}\
  \bibnamefont {{Poole}}},\ }\bibfield  {title} {\enquote {\bibinfo {title}
  {{Parallelized inference for gravitational-wave astronomy}},}\ }\href
  {\doibase 10.1103/PhysRevD.100.043030} {\bibfield  {journal} {\bibinfo
  {journal} {\prd}\ }\textbf {\bibinfo {volume} {100}},\ \bibinfo {eid}
  {043030} (\bibinfo {year} {2019})},\ \Eprint
  {http://arxiv.org/abs/1904.02863} {arXiv:1904.02863 [astro-ph.IM]}
  \BibitemShut {NoStop}%
\bibitem [{\citenamefont {Ashton}\ \emph {et~al.}(2019)\citenamefont {Ashton},
  \citenamefont {Hübner}, \citenamefont {Lasky}, \citenamefont {Talbot},
  \citenamefont {Ackley}, \citenamefont {Biscoveanu}, \citenamefont {Chu},
  \citenamefont {Divakarla}, \citenamefont {Easter}, \citenamefont {Goncharov},
  \citenamefont {Vivanco}, \citenamefont {Harms}, \citenamefont {Lower},
  \citenamefont {Meadors}, \citenamefont {Melchor}, \citenamefont {Payne},
  \citenamefont {Pitkin}, \citenamefont {Powell}, \citenamefont {Sarin},
  \citenamefont {Smith},\ and\ \citenamefont {Thrane}}]{2019_bilby}%
  \BibitemOpen
  \bibfield  {author} {\bibinfo {author} {\bibfnamefont {Gregory}\ \bibnamefont
  {Ashton}}, \bibinfo {author} {\bibfnamefont {Moritz}\ \bibnamefont
  {Hübner}}, \bibinfo {author} {\bibfnamefont {Paul~D.}\ \bibnamefont
  {Lasky}}, \bibinfo {author} {\bibfnamefont {Colm}\ \bibnamefont {Talbot}},
  \bibinfo {author} {\bibfnamefont {Kendall}\ \bibnamefont {Ackley}}, \bibinfo
  {author} {\bibfnamefont {Sylvia}\ \bibnamefont {Biscoveanu}}, \bibinfo
  {author} {\bibfnamefont {Qi}~\bibnamefont {Chu}}, \bibinfo {author}
  {\bibfnamefont {Atul}\ \bibnamefont {Divakarla}}, \bibinfo {author}
  {\bibfnamefont {Paul~J.}\ \bibnamefont {Easter}}, \bibinfo {author}
  {\bibfnamefont {Boris}\ \bibnamefont {Goncharov}}, \bibinfo {author}
  {\bibfnamefont {Francisco~Hernandez}\ \bibnamefont {Vivanco}}, \bibinfo
  {author} {\bibfnamefont {Jan}\ \bibnamefont {Harms}}, \bibinfo {author}
  {\bibfnamefont {Marcus~E.}\ \bibnamefont {Lower}}, \bibinfo {author}
  {\bibfnamefont {Grant~D.}\ \bibnamefont {Meadors}}, \bibinfo {author}
  {\bibfnamefont {Denyz}\ \bibnamefont {Melchor}}, \bibinfo {author}
  {\bibfnamefont {Ethan}\ \bibnamefont {Payne}}, \bibinfo {author}
  {\bibfnamefont {Matthew~D.}\ \bibnamefont {Pitkin}}, \bibinfo {author}
  {\bibfnamefont {Jade}\ \bibnamefont {Powell}}, \bibinfo {author}
  {\bibfnamefont {Nikhil}\ \bibnamefont {Sarin}}, \bibinfo {author}
  {\bibfnamefont {Rory J.~E.}\ \bibnamefont {Smith}}, \ and\ \bibinfo {author}
  {\bibfnamefont {Eric}\ \bibnamefont {Thrane}},\ }\bibfield  {title} {\enquote
  {\bibinfo {title} {Bilby: A user-friendly bayesian inference library for
  gravitational-wave astronomy},}\ }\href {\doibase 10.3847/1538-4365/ab06fc}
  {\bibfield  {journal} {\bibinfo  {journal} {The Astrophysical Journal
  Supplement Series}\ }\textbf {\bibinfo {volume} {241}},\ \bibinfo {pages}
  {27} (\bibinfo {year} {2019})}\BibitemShut {NoStop}%
\bibitem [{\citenamefont {Romero-Shaw}\ \emph {et~al.}(2020)\citenamefont
  {Romero-Shaw}, \citenamefont {Talbot}, \citenamefont {Biscoveanu},
  \citenamefont {D’Emilio}, \citenamefont {Ashton}, \citenamefont {Berry},
  \citenamefont {Coughlin}, \citenamefont {Galaudage}, \citenamefont {Hoy},
  \citenamefont {Hübner}, \citenamefont {Phukon}, \citenamefont {Pitkin},
  \citenamefont {Rizzo}, \citenamefont {Sarin}, \citenamefont {Smith},
  \citenamefont {Stevenson}, \citenamefont {Vajpeyi}, \citenamefont {Arène},
  \citenamefont {Athar}, \citenamefont {Banagiri}, \citenamefont {Bose},
  \citenamefont {Carney}, \citenamefont {Chatziioannou}, \citenamefont {Clark},
  \citenamefont {Colleoni}, \citenamefont {Cotesta}, \citenamefont {Edelman},
  \citenamefont {Estellés}, \citenamefont {García-Quirós}, \citenamefont
  {Ghosh}, \citenamefont {Green}, \citenamefont {Haster}, \citenamefont {Husa},
  \citenamefont {Keitel}, \citenamefont {Kim}, \citenamefont
  {Hernandez-Vivanco}, \citenamefont {Magaña Hernandez}, \citenamefont
  {Karathanasis}, \citenamefont {Lasky}, \citenamefont {De Lillo},
  \citenamefont {Lower}, \citenamefont {Macleod}, \citenamefont {Mateu-Lucena},
  \citenamefont {Miller}, \citenamefont {Millhouse}, \citenamefont {Morisaki},
  \citenamefont {Oh}, \citenamefont {Ossokine}, \citenamefont {Payne},
  \citenamefont {Powell}, \citenamefont {Pratten}, \citenamefont {Pürrer},
  \citenamefont {Ramos-Buades}, \citenamefont {Raymond}, \citenamefont
  {Thrane}, \citenamefont {Veitch}, \citenamefont {Williams}, \citenamefont
  {Williams},\ and\ \citenamefont {Xiao}}]{2020_bilby}%
  \BibitemOpen
  \bibfield  {author} {\bibinfo {author} {\bibfnamefont {I~M}\ \bibnamefont
  {Romero-Shaw}}, \bibinfo {author} {\bibfnamefont {C}~\bibnamefont {Talbot}},
  \bibinfo {author} {\bibfnamefont {S}~\bibnamefont {Biscoveanu}}, \bibinfo
  {author} {\bibfnamefont {V}~\bibnamefont {D’Emilio}}, \bibinfo {author}
  {\bibfnamefont {G}~\bibnamefont {Ashton}}, \bibinfo {author} {\bibfnamefont
  {C~P~L}\ \bibnamefont {Berry}}, \bibinfo {author} {\bibfnamefont
  {S}~\bibnamefont {Coughlin}}, \bibinfo {author} {\bibfnamefont
  {S}~\bibnamefont {Galaudage}}, \bibinfo {author} {\bibfnamefont
  {C}~\bibnamefont {Hoy}}, \bibinfo {author} {\bibfnamefont {M}~\bibnamefont
  {Hübner}}, \bibinfo {author} {\bibfnamefont {K~S}\ \bibnamefont {Phukon}},
  \bibinfo {author} {\bibfnamefont {M}~\bibnamefont {Pitkin}}, \bibinfo
  {author} {\bibfnamefont {M}~\bibnamefont {Rizzo}}, \bibinfo {author}
  {\bibfnamefont {N}~\bibnamefont {Sarin}}, \bibinfo {author} {\bibfnamefont
  {R}~\bibnamefont {Smith}}, \bibinfo {author} {\bibfnamefont {S}~\bibnamefont
  {Stevenson}}, \bibinfo {author} {\bibfnamefont {A}~\bibnamefont {Vajpeyi}},
  \bibinfo {author} {\bibfnamefont {M}~\bibnamefont {Arène}}, \bibinfo
  {author} {\bibfnamefont {K}~\bibnamefont {Athar}}, \bibinfo {author}
  {\bibfnamefont {S}~\bibnamefont {Banagiri}}, \bibinfo {author} {\bibfnamefont
  {N}~\bibnamefont {Bose}}, \bibinfo {author} {\bibfnamefont {M}~\bibnamefont
  {Carney}}, \bibinfo {author} {\bibfnamefont {K}~\bibnamefont
  {Chatziioannou}}, \bibinfo {author} {\bibfnamefont {J~A}\ \bibnamefont
  {Clark}}, \bibinfo {author} {\bibfnamefont {M}~\bibnamefont {Colleoni}},
  \bibinfo {author} {\bibfnamefont {R}~\bibnamefont {Cotesta}}, \bibinfo
  {author} {\bibfnamefont {B}~\bibnamefont {Edelman}}, \bibinfo {author}
  {\bibfnamefont {H}~\bibnamefont {Estellés}}, \bibinfo {author}
  {\bibfnamefont {C}~\bibnamefont {García-Quirós}}, \bibinfo {author}
  {\bibfnamefont {Abhirup}\ \bibnamefont {Ghosh}}, \bibinfo {author}
  {\bibfnamefont {R}~\bibnamefont {Green}}, \bibinfo {author} {\bibfnamefont
  {C-J}\ \bibnamefont {Haster}}, \bibinfo {author} {\bibfnamefont
  {S}~\bibnamefont {Husa}}, \bibinfo {author} {\bibfnamefont {D}~\bibnamefont
  {Keitel}}, \bibinfo {author} {\bibfnamefont {A~X}\ \bibnamefont {Kim}},
  \bibinfo {author} {\bibfnamefont {F}~\bibnamefont {Hernandez-Vivanco}},
  \bibinfo {author} {\bibfnamefont {I}~\bibnamefont {Magaña Hernandez}},
  \bibinfo {author} {\bibfnamefont {C}~\bibnamefont {Karathanasis}}, \bibinfo
  {author} {\bibfnamefont {P~D}\ \bibnamefont {Lasky}}, \bibinfo {author}
  {\bibfnamefont {N}~\bibnamefont {De Lillo}}, \bibinfo {author}
  {\bibfnamefont {M~E}\ \bibnamefont {Lower}}, \bibinfo {author} {\bibfnamefont
  {D}~\bibnamefont {Macleod}}, \bibinfo {author} {\bibfnamefont
  {M}~\bibnamefont {Mateu-Lucena}}, \bibinfo {author} {\bibfnamefont
  {A}~\bibnamefont {Miller}}, \bibinfo {author} {\bibfnamefont {M}~\bibnamefont
  {Millhouse}}, \bibinfo {author} {\bibfnamefont {S}~\bibnamefont {Morisaki}},
  \bibinfo {author} {\bibfnamefont {S~H}\ \bibnamefont {Oh}}, \bibinfo {author}
  {\bibfnamefont {S}~\bibnamefont {Ossokine}}, \bibinfo {author} {\bibfnamefont
  {E}~\bibnamefont {Payne}}, \bibinfo {author} {\bibfnamefont {J}~\bibnamefont
  {Powell}}, \bibinfo {author} {\bibfnamefont {G}~\bibnamefont {Pratten}},
  \bibinfo {author} {\bibfnamefont {M}~\bibnamefont {Pürrer}}, \bibinfo
  {author} {\bibfnamefont {A}~\bibnamefont {Ramos-Buades}}, \bibinfo {author}
  {\bibfnamefont {V}~\bibnamefont {Raymond}}, \bibinfo {author} {\bibfnamefont
  {E}~\bibnamefont {Thrane}}, \bibinfo {author} {\bibfnamefont {J}~\bibnamefont
  {Veitch}}, \bibinfo {author} {\bibfnamefont {D}~\bibnamefont {Williams}},
  \bibinfo {author} {\bibfnamefont {M~J}\ \bibnamefont {Williams}}, \ and\
  \bibinfo {author} {\bibfnamefont {L}~\bibnamefont {Xiao}},\ }\bibfield
  {title} {\enquote {\bibinfo {title} {Bayesian inference for compact binary
  coalescences with bilby: validation and application to the first ligo–virgo
  gravitational-wave transient catalogue},}\ }\href {\doibase
  10.1093/mnras/staa2850} {\bibfield  {journal} {\bibinfo  {journal} {Monthly
  Notices of the Royal Astronomical Society}\ }\textbf {\bibinfo {volume}
  {499}},\ \bibinfo {pages} {3295–3319} (\bibinfo {year} {2020})}\BibitemShut
  {NoStop}%
\bibitem [{\citenamefont {Thrane}\ and\ \citenamefont
  {Talbot}(2019)}]{2019_Bayesian}%
  \BibitemOpen
  \bibfield  {author} {\bibinfo {author} {\bibfnamefont {Eric}\ \bibnamefont
  {Thrane}}\ and\ \bibinfo {author} {\bibfnamefont {Colm}\ \bibnamefont
  {Talbot}},\ }\bibfield  {title} {\enquote {\bibinfo {title} {An introduction
  to bayesian inference in gravitational-wave astronomy: Parameter estimation,
  model selection, and hierarchical models},}\ }\href {\doibase
  10.1017/pasa.2019.2} {\bibfield  {journal} {\bibinfo  {journal} {Publications
  of the Astronomical Society of Australia}\ }\textbf {\bibinfo {volume} {36}}
  (\bibinfo {year} {2019}),\ 10.1017/pasa.2019.2}\BibitemShut {NoStop}%
\bibitem [{\citenamefont {Pratten}\ \emph {et~al.}(2021)\citenamefont
  {Pratten}, \citenamefont {García-Quirós}, \citenamefont {Colleoni},
  \citenamefont {Ramos-Buades}, \citenamefont {Estellés}, \citenamefont
  {Mateu-Lucena}, \citenamefont {Jaume}, \citenamefont {Haney}, \citenamefont
  {Keitel}, \citenamefont {Thompson},\ and\ \citenamefont
  {Husa}}]{2021_IMRPhenomXPHM}%
  \BibitemOpen
  \bibfield  {author} {\bibinfo {author} {\bibfnamefont {Geraint}\ \bibnamefont
  {Pratten}}, \bibinfo {author} {\bibfnamefont {Cecilio}\ \bibnamefont
  {García-Quirós}}, \bibinfo {author} {\bibfnamefont {Marta}\ \bibnamefont
  {Colleoni}}, \bibinfo {author} {\bibfnamefont {Antoni}\ \bibnamefont
  {Ramos-Buades}}, \bibinfo {author} {\bibfnamefont {Héctor}\ \bibnamefont
  {Estellés}}, \bibinfo {author} {\bibfnamefont {Maite}\ \bibnamefont
  {Mateu-Lucena}}, \bibinfo {author} {\bibfnamefont {Rafel}\ \bibnamefont
  {Jaume}}, \bibinfo {author} {\bibfnamefont {Maria}\ \bibnamefont {Haney}},
  \bibinfo {author} {\bibfnamefont {David}\ \bibnamefont {Keitel}}, \bibinfo
  {author} {\bibfnamefont {Jonathan~E.}\ \bibnamefont {Thompson}}, \ and\
  \bibinfo {author} {\bibfnamefont {Sascha}\ \bibnamefont {Husa}},\ }\bibfield
  {title} {\enquote {\bibinfo {title} {Computationally efficient models for the
  dominant and subdominant harmonic modes of precessing binary black holes},}\
  }\href {\doibase 10.1103/physrevd.103.104056} {\bibfield  {journal} {\bibinfo
   {journal} {Physical Review D}\ }\textbf {\bibinfo {volume} {103}} (\bibinfo
  {year} {2021}),\ 10.1103/physrevd.103.104056}\BibitemShut {NoStop}%
\bibitem [{Note4()}]{Note4}%
  \BibitemOpen
  \bibinfo {note} {The updated version of Ref.~\cite {2021_better_spin}
  corrects for another bug, which resulted in the non-spinning posteriors to be
  given more weight where the fiducial prior per event was 4 times larger than
  it should have been. However, the update to \cite {2021_better_spin} does not
  address the suspect samples for $\protect \rm {GW}190408\protect \_181802$,
  which we fix here.}\BibitemShut {Stop}%
\bibitem [{Note5()}]{Note5}%
  \BibitemOpen
  \bibinfo {note} {Supplementary material including hyper-posterior samples for
  all model variants and result plots are available here: \protect \url
  {https://github.com/HuiTong5/GWTC-3_pop_spin}}\BibitemShut {NoStop}%
\bibitem [{\citenamefont {Miller}\ \emph {et~al.}(2020)\citenamefont {Miller},
  \citenamefont {Callister},\ and\ \citenamefont {Farr}}]{Miller2020}%
  \BibitemOpen
  \bibfield  {author} {\bibinfo {author} {\bibfnamefont {Simona}\ \bibnamefont
  {Miller}}, \bibinfo {author} {\bibfnamefont {Thomas~A.}\ \bibnamefont
  {Callister}}, \ and\ \bibinfo {author} {\bibfnamefont {Will~M.}\ \bibnamefont
  {Farr}},\ }\bibfield  {title} {\enquote {\bibinfo {title} {The low effective
  spin of binary black holes and implications for individual gravitational-wave
  events},}\ }\href@noop {} {\bibfield  {journal} {\bibinfo  {journal}
  {Astrophys. J.}\ }\textbf {\bibinfo {volume} {895}},\ \bibinfo {pages} {128}
  (\bibinfo {year} {2020})}\BibitemShut {NoStop}%
\bibitem [{\citenamefont {Callister}\ \emph
  {et~al.}(2021{\natexlab{b}})\citenamefont {Callister}, \citenamefont
  {Haster}, \citenamefont {Ng}, \citenamefont {Vitale},\ and\ \citenamefont
  {Farr}}]{Callister2021}%
  \BibitemOpen
  \bibfield  {author} {\bibinfo {author} {\bibfnamefont {Thomas~A.}\
  \bibnamefont {Callister}}, \bibinfo {author} {\bibfnamefont {Carl-Johan}\
  \bibnamefont {Haster}}, \bibinfo {author} {\bibfnamefont {Ken K.~Y.}\
  \bibnamefont {Ng}}, \bibinfo {author} {\bibfnamefont {Salvatore}\
  \bibnamefont {Vitale}}, \ and\ \bibinfo {author} {\bibfnamefont {Will~M.}\
  \bibnamefont {Farr}},\ }\bibfield  {title} {\enquote {\bibinfo {title} {{Who
  Ordered That? Unequal-mass Binary Black Hole Mergers Have Larger Effective
  Spins}},}\ }\href@noop {} {\bibfield  {journal} {\bibinfo  {journal}
  {Astrophys. J. Lett.}\ }\textbf {\bibinfo {volume} {922}} (\bibinfo {year}
  {2021}{\natexlab{b}})}\BibitemShut {NoStop}%
\bibitem [{\citenamefont {Broekgaarden}\ \emph {et~al.}(2022)\citenamefont
  {Broekgaarden}, \citenamefont {Stevenson},\ and\ \citenamefont
  {Thrane}}]{Broekgaarden2022}%
  \BibitemOpen
  \bibfield  {author} {\bibinfo {author} {\bibfnamefont {F}~\bibnamefont
  {Broekgaarden}}, \bibinfo {author} {\bibfnamefont {S}~\bibnamefont
  {Stevenson}}, \ and\ \bibinfo {author} {\bibfnamefont {E}~\bibnamefont
  {Thrane}},\ }\bibfield  {title} {\enquote {\bibinfo {title} {Signatures of
  mass ratio reversal in gravitational waves from merging binary black
  holes},}\ }\href@noop {} {\  (\bibinfo {year} {2022})},\ \bibinfo {note}
  {arxiv/2205.01693}\BibitemShut {NoStop}%
\bibitem [{\citenamefont {Qin}\ \emph {et~al.}(2022)\citenamefont {Qin},
  \citenamefont {Wang}, \citenamefont {Wu}, \citenamefont {Meynet},\ and\
  \citenamefont {Song}}]{Qin2022}%
  \BibitemOpen
  \bibfield  {author} {\bibinfo {author} {\bibfnamefont {Ying}\ \bibnamefont
  {Qin}}, \bibinfo {author} {\bibfnamefont {Yuan-Zhu}\ \bibnamefont {Wang}},
  \bibinfo {author} {\bibfnamefont {Dong-Hong}\ \bibnamefont {Wu}}, \bibinfo
  {author} {\bibfnamefont {Georges}\ \bibnamefont {Meynet}}, \ and\ \bibinfo
  {author} {\bibfnamefont {Hanfeng}\ \bibnamefont {Song}},\ }\bibfield  {title}
  {\enquote {\bibinfo {title} {On the angular momentum transport efficiency
  within the star constrained from gravitational-wave observations},}\
  }\href@noop {} {\bibfield  {journal} {\bibinfo  {journal} {Astrophys. J.}\
  }\textbf {\bibinfo {volume} {924}},\ \bibinfo {pages} {129} (\bibinfo {year}
  {2022})}\BibitemShut {NoStop}%
\bibitem [{\citenamefont {Mandel}\ and\ \citenamefont
  {Fragos}(2020)}]{Mandel2020}%
  \BibitemOpen
  \bibfield  {author} {\bibinfo {author} {\bibfnamefont {Ilya}\ \bibnamefont
  {Mandel}}\ and\ \bibinfo {author} {\bibfnamefont {Tassos}\ \bibnamefont
  {Fragos}},\ }\bibfield  {title} {\enquote {\bibinfo {title} {{An alternative
  interpretation of GW190412 as a binary black hole merger with a rapidly
  spinning secondary}},}\ }\href@noop {} {\bibfield  {journal} {\bibinfo
  {journal} {Astrophys. J. Lett.}\ }\textbf {\bibinfo {volume} {895}},\
  \bibinfo {pages} {L28} (\bibinfo {year} {2020})}\BibitemShut {NoStop}%
\bibitem [{\citenamefont {Biscoveanu}\ \emph {et~al.}(2021)\citenamefont
  {Biscoveanu}, \citenamefont {Isi}, \citenamefont {Vitale},\ and\
  \citenamefont {Varma}}]{Biscoveanu_2021}%
  \BibitemOpen
  \bibfield  {author} {\bibinfo {author} {\bibfnamefont {Sylvia}\ \bibnamefont
  {Biscoveanu}}, \bibinfo {author} {\bibfnamefont {Maximiliano}\ \bibnamefont
  {Isi}}, \bibinfo {author} {\bibfnamefont {Salvatore}\ \bibnamefont {Vitale}},
  \ and\ \bibinfo {author} {\bibfnamefont {Vijay}\ \bibnamefont {Varma}},\
  }\bibfield  {title} {\enquote {\bibinfo {title} {New spin on {LIGO}-virgo
  binary black holes},}\ }\href {\doibase 10.1103/physrevlett.126.171103}
  {\bibfield  {journal} {\bibinfo  {journal} {Physical Review Letters}\
  }\textbf {\bibinfo {volume} {126}} (\bibinfo {year} {2021}),\
  10.1103/physrevlett.126.171103}\BibitemShut {NoStop}%
\end{thebibliography}%


%

\appendix
\section{Additional results}

\begin{figure*}[htbp!]
    \centering
    \includegraphics[width=2\columnwidth]{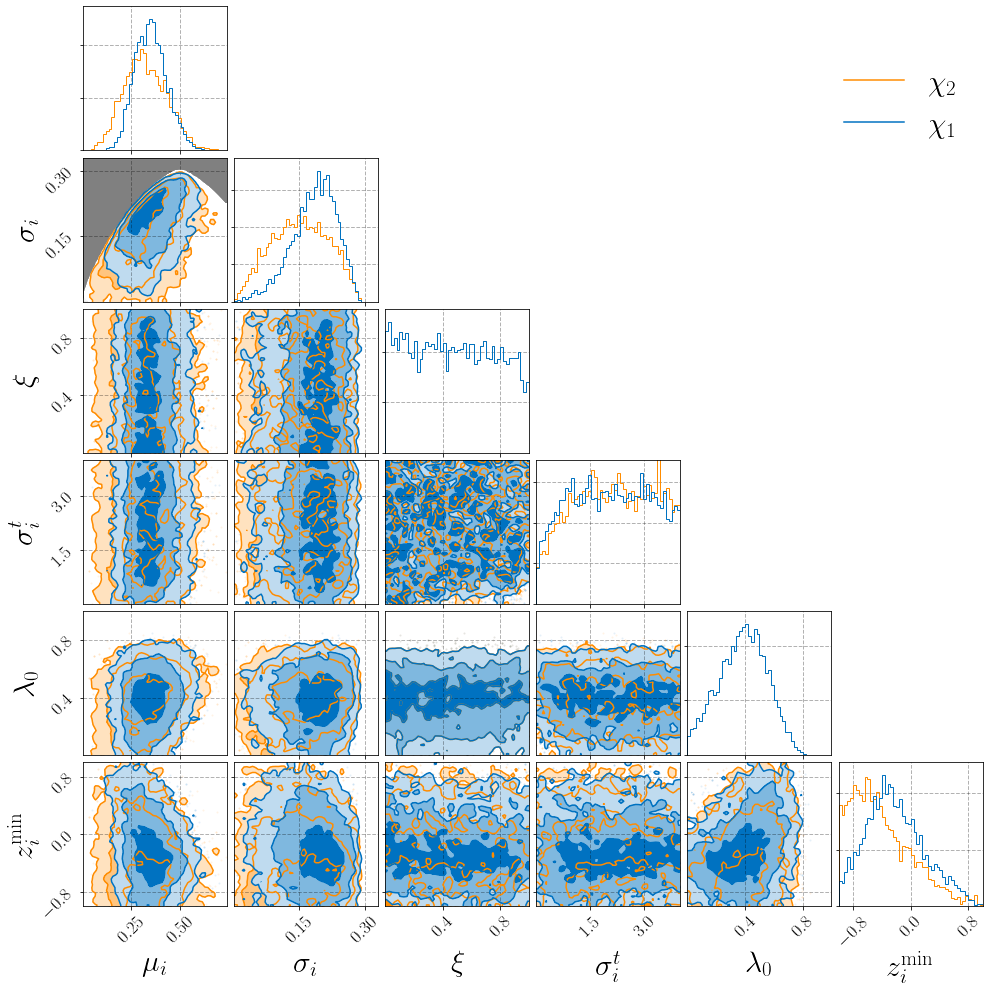}
    \caption{
    A corner plot showing the population parameters from the best-fit \nonid model variant. (GW191109 is excluded.)
    The results for $\chi_1$ are shown in blue while the results from $\chi_2$ are in orange. We marked the forbidden region in $(\mu_i,\sigma_i)$ panel. It is a restriction arising from the positivity of dimensionless spin magnitude $\chi$.
    }
    \label{fig:corner_overplot}
\end{figure*}

\begin{figure}
    \centering
    \includegraphics[width=1\columnwidth]{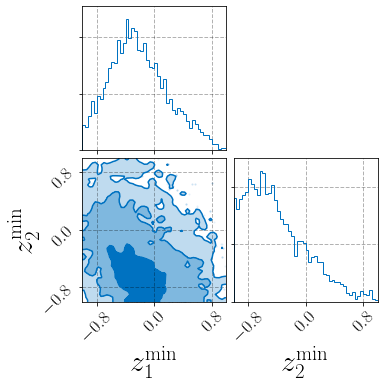}
    \caption{A corner plot showing hyper-parameters $z^{min}_1$ versus $z^{min}_2$ from \nonid model variant. (GW191109 is excluded.)}
    \label{fig:ind_zmin}
\end{figure}

\begin{figure*}
    \centering
    \includegraphics[width=2\columnwidth]{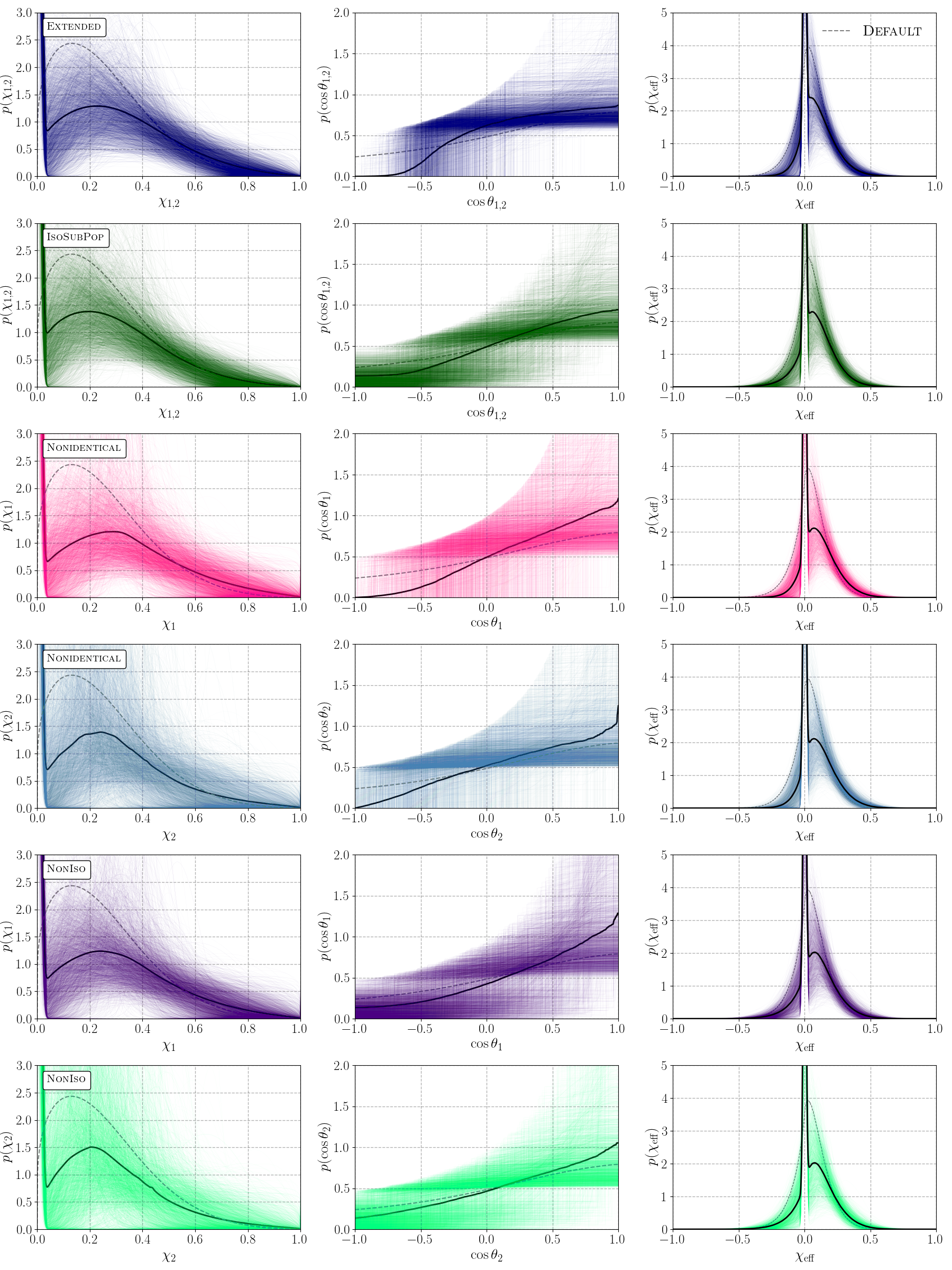}
    \caption{Population predictive distributions for dimensionless spin $\chi$, cosine tilt angle $z$ and effective inspiral spin $\chi_\text{eff}$ given  different model variants. (GW191109 is excluded.) For model variants with nonidentical $\chi_1,2$, the PPD for $\chi_{\text{eff}}$ is the same.}
    \label{fig:all_ppd}
\end{figure*}

\begin{figure*}[htbp!]
    \begin{subfigure}
    \centering
    \includegraphics[width=0.49\linewidth]{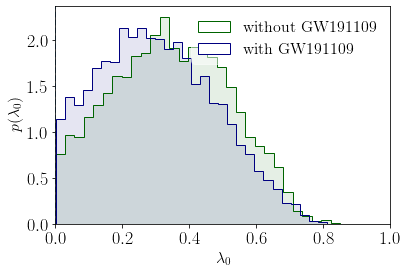}
    \end{subfigure}
    \hfill
    \begin{subfigure}
    \centering
    \includegraphics[width=0.49\linewidth]{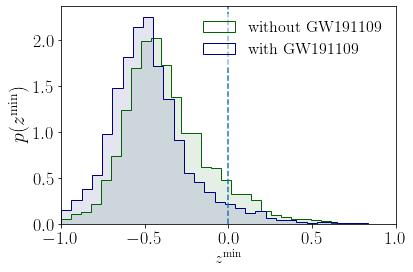}
    \end{subfigure}
    \caption{The posterior distributions for $z^\text{min}$ and $\lambda_0$ using \exdefault model. The colors denote the dataset (GWTC-3 with and without the potentially problematic event, GW191109).}
    \label{fig:zmin_lam_191109}
\end{figure*}

\begin{figure}
    \centering
    \includegraphics[width=\columnwidth]{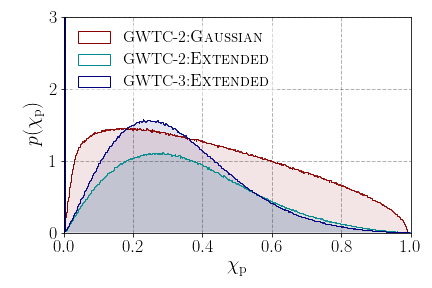}
    \caption{Population predictive reconstructed distribution for $\chi_\text{p}$ for the \gaussian model and the \exdefault model using GWTC-2/3 data. We exclude GW191109.}
    \label{fig:chi_p}
\end{figure}

This appendix includes additional material, which may be of use to experts in gravitational-wave astronomy.
\begin{itemize}
    \item In Fig.~\ref{fig:corner_overplot}, we provide a corner plot showing the posteriors of all hyper-parameters related to spin properties in \nonid model. In addition, we separately show a corner plot in Fig.~\ref{fig:ind_zmin} for $z^{min}_1$ versus $z^{min}_2$. $z_1^{min} >= 0$ and $z_2^{min} >= 0$ are allowed simultaneously as an evidence for isolated binary isolation. 
    \item In Table~\ref{tab:hyper_parameters}, we summarize the median and 90\% credible intervals for  key hyper-parameters.
    \item In Fig.~\ref{fig:all_ppd}, we show PPD plots for dimensionless spin $\chi$, cosine tilt angle $z$ and effective inspiral spin $\chi_\text{eff}$ given all different model variants.
    \item In Fig.~\ref{fig:zmin_lam_191109}, we show posteriors for key population hyper-parameters obtained while including the event GW191109.
    
    \item In Fig.~\ref{fig:chi_p}, we show the PPD plot for the effective precession spin parameter $\chi_\text{p}$ of \gaussian and \exdefault model using GWTC-2/3 data.
\end{itemize}

\setlength{\tabcolsep}{18pt}
\renewcommand{\arraystretch}{1.5}
\begin{table*}
\centering
 \begin{tabular}{p{4cm} | c | c| c | c | c   } 
 \hline
 Model & $z^\text{min}_{1}$& $z^\text{min}_2$ & $\lambda_0$ & $\zeta_{99\%}$ & $\zeta_{1\%}$ \\
 \hline\hline
 \nonid & $-0.26_{-0.46}^{+0.62}$ &$-0.49_{-0.40}^{+0.68}$ & $0.39_{-0.24}^{+0.20}$& 0.09 & 0.99 \\
 \exdefault& $-0.41_{-0.23}^{+0.39}$ & - & $-0.34_{-0.23}^{+0.24}$ & 0.10 & 0.99 \\
 \isp& $-0.23_{-0.42}^{+0.57}$& - & $-0.34_{-0.23}^{+0.22}$ & 0.45 & 1.00 \\
 \nonid \isp & $-0.15_{-0.58}^{+0.69}$ & $-0.33_{-0.54}^{+0.74}$ & $0.38_{-0.22}^{+0.20}$ & 0.44 & 1.00 \\
 \hline
 \nonid with $\lambda_0=0$ & $-0.42_{-0.32}^{+0.31}$ & $-0.63_{-0.29}^{+0.46}$ & 0 & 0.07 & 0.98\\
 \exdefault with $\lambda_0=0$& $-0.51_{-0.18}^{+0.14}$ & - & 0 &  0.09& 0.99 \\
 \exdefault with $z^\text{min}=-1$ & $-1$ & - & $0.34_{-0.22}^{+0.21}$ & 0.46 & 1.00\\
 \hline
 \default & $-1$ & - & 0 & 0.42 & 1.00\\
 \hline
 \end{tabular}
 \caption{Median and 90\% credible intervals on various hyper-parameters in our models. GW191109 is excluded in these analyses.
 The $z_\text{min}$ parameter(s) determine the minimum value of the cosine of the black-hole spin vector with respect to the orbital angular momentum axis.
 The parameter $\lambda_0$ is the fraction of BBH mergers with zero black-hole spin.
 The last two columns provide the $(1\%,99\%)$ credible interval for $\zeta$, the fraction of ``field-like'' binaries (with preferentially aligned spins).
 }
 \label{tab:hyper_parameters}
\end{table*}

\begin{table*}
\centering
 \begin{tabular}{p{0.7cm} p{10cm} p{1.2cm}} 
 \hline
 \textbf{Parameter} & \textbf{Description} & \textbf{Prior}\\
 \hline\hline
 $\lambda_0$ & Mixing fraction of mergers with zero spin, $\chi_1=\chi_2=0$ & U(0,1)\\
 $\mu_i$ & Mean of spin magnitude distribution & U(0,1)\\
 $\sigma^2_i$ & The square of the width of the spin magnitude distribution & U(0,0.25)\\
 $\zeta$ & Mixing fraction of mergers with preferentially aligned spin & U(0,1)\\
 $\sigma^t_i$ & Spread in projected misalignment for preferentially aligned black holes & U(0,4)\\
 $z^{min}_i$ & Minimum value of the projected misalignment & U($-1$,1)\\
 \hline
 \end{tabular}
 \caption{A summary of priors for population hyper-parameters. The notation $U(a,b)$ indicates a uniform distribution on the interval $(a,b)$.}
 \label{tab:priors}
\end{table*}

\end{document}